\documentclass[aps,prx,reprint,floatfix,groupeaddress,showkeys,longbibliography]{revtex4-1}	
\usepackage[utf8]{inputenc}     
\usepackage[english]{babel}   

\usepackage{lipsum}
\usepackage[hidelinks,colorlinks=true]{hyperref}

\usepackage[usenames,dvipsnames,svgnames,table]{xcolor}
\usepackage{graphicx}

\definecolor{myred}{RGB}{228,26,28}
\definecolor{myblue}{RGB}{55,126,184}
\definecolor{myorange}{RGB}{225,127,0}
\definecolor{mygreen}{RGB}{77,175,74}
\definecolor{mylila}{RGB}{152,78,163}
\definecolor{mybrown}{RGB}{153,76,0}
\definecolor{mygray}{RGB}{153,153,153}
\definecolor{darkred}{rgb}{0.8,0,0}
\definecolor{mydarkgreen}{RGB}{0,102,0}
\definecolor{mydarkbrown}{RGB}{102,52,0}
\definecolor{Orange}{RGB}{235,129,27}
\definecolor{Green}{RGB}{35,55,59}
\usepackage{bm}
\usepackage{amsmath,amsfonts,amssymb}
\newcommand{\comment}[1]{\textcolor{black}{#1}}
\newcommand{\newcomment}[1]{\textcolor{black}{#1}}
\newcommand{\idest}{\emph{i.e. }}

\newcommand{\const}{\operatorname{ const} }

\renewcommand{\d}{\mathrm{d}}
\newcommand{\D}{\partial}

\newcommand{\euler}{\mathrm{e}}
\renewcommand{\i}{\mathrm{i}}

\newcommand{\x}{\bm{x}}
\newcommand{\y}{\bm{y}}

\newcommand{\W}{\bm{W}}

\newcommand{\Real}{\operatorname{Re}}
\newcommand{\Imag}{\operatorname{Im}}

\newcommand{\invtemperature}{\beta}

\newcommand{\force}{f}
\newcommand{\maxforce}{f^*}
\newcommand{\inforce}{f_1}

\newcommand{\outforce}{f_2}
\newcommand{\maxoutforce}{f_2^*}
\newcommand{\microsign}{\Theta(\microstate,\microstate')}
\newcommand{\macrosign}{\Theta(\macrostate,\macrostate')}
\newcommand{\meanfieldsign}{\Theta(i,j)}
\newcommand{\change}{\Delta}
\newcommand{\boltzmann}{k_b}
\newcommand{\Potential}{U}
\newcommand{\potential}{u}
\newcommand{\dimension}{N}
\newcommand{\occupation}{N}
\newcommand{\occupationdensity}{n}
\newcommand{\stateenergy}{\epsilon}
\newcommand{\microstate}{\alpha}
\newcommand{\macrostate}{\bm{N}}
\newcommand{\arrheniusprefactor}{\Gamma}

\newcommand{\meanfieldeigenvalue}{\lambda}
\newcommand{\masterequationeigenvalue}{\lambda}
\newcommand{\jacobian}{\bm{J}}
\newcommand{\complexvar}{z}

\newcommand{\lefteigenvectors}{\Phi^L}
\newcommand{\righteigenvectors}{\Phi^R}
\newcommand{\limitcyclefrequency}{\omega_{lc} }

\newcommand{\linearstabilityfrequency}{\omega_{ls}}

\newcommand{\jointprobability}{\mathfrak{P}}
\newcommand{\microprobability}{p}
\newcommand{\microsteadyprobability}{p^{s}}
\newcommand{\microeqprobability}{p^{eq}}
\newcommand{\macroprobability}{P}
\newcommand{\macrosteadyprobability}{P^{s}}
\newcommand{\macroeqprobability}{P^{eq}}
\newcommand{\meanfieldprobability}{\overline{n}}
\newcommand{\meanfieldprobabilityvector}{\bm{\overline{n}}}
\newcommand{\meanfieldsteadyprobability}{\overline{n}^s}
\newcommand{\meanfieldeqprobability}{\overline{n}^{eq}}

\newcommand{\macroprobabilityvector}{\bm{P}}

\newcommand{\multiplicity}{\Omega}
\newcommand{\microrates}{w}
\newcommand{\macrorates}{W}
\newcommand{\marginalizedrates}{\tilde{\macrorates}}
\newcommand{\rates}{W}
\newcommand{\meanfieldrates}{k}

\newcommand{\macroratematrix}{\bm{W}}

\newcommand{\tree}{\mathcal{T}}
\newcommand{\graph}{G}

\newcommand{\microenergy}{e}
\newcommand{\microheat}{q}
\newcommand{\microwork}{w}
\newcommand{\microentropy}{s}

\newcommand{\microep}{\sigma}

\newcommand{\macroenergy}{E}
\newcommand{\macroheat}{Q}
\newcommand{\macrowork}{W}
\newcommand{\macroentropy}{S}
\newcommand{\macroentropyflow}{S_e}
\newcommand{\macroep}{\Sigma}
\newcommand{\internalentropy}{S^{int}}
\newcommand{\microfreeenergy}{A}
\newcommand{\macrofreeenergy}{A}

\newcommand{\meanfieldenergy}{\mathcal{E}}
\newcommand{\meanfieldheat}{\mathcal{Q}}
\newcommand{\meanfieldwork}{\mathcal{W}}
\newcommand{\meanfieldpower}{\mathcal{P}}
\newcommand{\maxmeanfieldpower}{\mathcal{P}^*}
\newcommand{\meanfieldentropy}{\mathcal{S}}

\newcommand{\meanfieldep}{\mathcal{S}_i}

\newcommand{\singleworkrate}{ \overline{\macrowork}_{1} }
\newcommand{\singleworkdifference}{ \Delta \overline{\macrowork}_{1 \dimension} }
\newcommand{\finiteworkdifference}{ \Delta \overline{\macrowork}_{\dimension,10^4} }

\newcommand{\efficiency}{\eta}
\newcommand{\maxefficiency}{\eta^*}

\begin{document}
\title{Collective power:\\Minimal model for thermodynamics of nonequilibrium phase transitions}
%\title{The power of working together:\\Minimal model for thermodynamics of nonequilibrium phase transitions}
%\title{Synchronization works:\\Minimal model for thermodynamics of nonequilibrium phase transitions}
%\title{Collective power:\\Minimal model for thermodynamics of nonequilibrium phase transitions}
\author{Tim Herpich}
\email{Electronic Mail: tim.herpich@uni.lu}
\author{Juzar Thingna}
\author{Massimiliano Esposito}
\affiliation{Complex Systems and Statistical Mechanics, Physics and Materials Science Research Unit, University of Luxembourg, L-1511 Luxembourg, Luxembourg}
\date{\today}

\begin{abstract}
%We establish a direct connection between the linear stochastic dynamics, the nonlinear mean-field dynamics, and the thermodynamic description of a minimal model of driven and interacting \comment{three-state units}. 
\comment{We propose a thermodynamically consistent minimal model to study synchronization which is made of driven and interacting three-state units.
This system exhibits at the mean-field level two bifurcations separating three dynamical phases: a single stable fixed point, a stable limit cycle indicative of synchronization, and multiple stable fixed points.
These complex emergent dynamical behaviors are understood at the level of the underlying linear Markovian dynamics in terms of metastability, i.e. the appearance of gaps in the upper real part of the spectrum of the Markov generator.
%The apparent contradiction with the underlying linear Markovian dynamics which ensures convergence to a unique steady state is resolved via metastability, i.e. the appearance of gaps in the upper real part of the spectrum of the Markov generator.
Stochastic thermodynamics is used to study the dissipated work across dynamical phases as well as across scales. 
%Thermodynamically, the dissipated work of the stochastic dynamics exhibits signatures of nonequilibrium phase transitions over long metastable times which disappear in the infinite-time limit. 
This dissipated work is found to be reduced by the attractive interactions between the units and to nontrivially depend on the system size.
When operating as a work-to-work converter, we find that the maximum power output is achieved far-from-equilibrium in the synchronization regime and that the efficiency at maximum power is surprisingly close to the linear regime prediction.
Our work shows the way towards building a thermodynamics of nonequilibrium phase transitions in conjunction to bifurcation theory.
%Stochastic thermodynamics and bifurcation theory have a priori little in common. The former is built upon linear Markovian stochastic dynamics while the latter studies topological changes in nonlinear dynamical systems. 
}
\end{abstract}
\maketitle

\section{Introduction}

%{\bf check oscillators vs units, add ref of Ruffo}

While phase transitions are quite well understood at equilibrium, nonequilibrium phase transitions still lack a systematic treatment. They are most commonly described as dynamical phenomenon within the framework of nonlinear dynamics and bifurcation theory \cite{Strogatz, Guckenheimer1983}, but their relation to thermodynamics is rarely discussed. This is largely due to the fact that a theory of nonequilibrium thermodynamics was lacking. Stochastic thermodynamics nowadays provides one for systems described by stochastic dynamics \cite{seifert2012rpp, broeck2015physica, zhang2012pr, rao2018njp}. But until recently it has been mostly explored to study noninteracting systems or systems made of few interacting particles. We will use stochastic thermodynamics to explore the physics of nonequilibrium phase transitions in large ensembles of interacting systems.

A motivation to do so which is of great practical importance is to understand how phase transitions, and more generally interactions, affect the performance of large ensembles of nano-machines.
Indeed, while these latter have been shown to make very good energy converters, the main drawback remains their low power output. A natural way out is to assemble large numbers of nano-machines, which immediately raises the question of whether certain interactions are favorable to their overall performance. Stochastic thermodynamics provides a powerful framework to do so as it has proved instrumental to analyze the performance of small energy converters operating far-from-equilibrium \cite{seifert2012rpp, pekola2015np, ciliberto2017prx} (e.g. thermoelectric quantum dots \cite{esposito2009epl, bulnes2015njp}, photoelectric nanocells \cite{esposito2009prb}, molecular motors \cite{ge2012pr, gaspard2007jtb, lacoste2007prl, seifert2011epj, altaner2015pre}) and their power-efficiency trade-off \cite{shiraishi2016prl, proesmans2016prl, pietzonka2017arxiv, park2017srp, polettini2017epl}. We emphasize that going beyond linear response is essential here since this is where nonequilibrium phase transitions occur. While some works have been done in this direction, most are restricted to mean-field treatments \cite{verley2017epl, shiraishi2017pre, imparato2012prl, imparato2012prl, imparato2012pre, imparato2015njp, sasa2015njp}. An important aspect of our study will be to analyze in details the emergence of the mean-field description from the underlying stochastic dynamics.

The paradigmatic phase transition which we will consider is synchronization: coupled units with different natural frequencies exhibiting a spontaneous phase-locking to a global frequency \cite{pikovsky2003}. This collective phenomenon was famously described by Huygens \cite{huygens} who experimentally observed that two pendulum clocks attached to a common support display an ``odd kind of sympathy'' \cite{huygens}, that is they synchronize in anti-phase. It was later found to be ubiquitous in nature \cite{strogatz2004}. 
%common example being the synchronous flashing of fireflies, or the swimming and flying patterns of fish and birds. 
Synchronization is typically modeled by coupled phase oscillators which exhibit phase-locking when the coupling strength exceeds a critical value \cite{winfree1967jtb}. The most commonly used (noisy) Kuramoto model \cite{kuramoto2003, strogatz2000pd, aceborn2005rmp, RuffoJSTAT2014} is well understood for an infinite population of oscillator at the mean-field level. Some works also considered few locally coupled oscillators \cite{sakaguchi1987ptp, daido1988prl, walgraef1976jcp} and even the dissipation resulting from their synchronization \cite{seifert2016pre}. However, little is known about large but finite populations of stochastic oscillators (see e.g. Refs. \cite{risler2004prl, risler2005pre}). Progress in this direction was done in Refs. \cite{wood2006prl, wood2006pre, wood2007pre} by introducing a minimal stochastic model made of interacting three-state units and shown to exhibit phase synchronization. It enabled to compare the mean-field dynamics to the Monte-Carlo one. \comment{However, since this model is made of three unidirectional stochastic transitions, it does not allow for a consistent thermodynamic description. Furthermore, the extent in which this ingredient is essential for synchronization is not clear.} These works also did not provide a detailed understanding of how a linear and irreducible Markov dynamics can give rise to a nonlinear mean-field dynamics with increasing system size. This question is particularly intriguing since the Perron-Frobenius theorem ensures that the former dynamics has a unique stationary solution (for finite state spaces) \cite{vankampen2007} while the latter can exhibit multiple and time-periodic solutions. \comment{It is also closely related to the emergence of hydrodynamic modes or metastability \cite{GaspardPRE96, GaspardB, GaveauSchulmanJPA98, KurchanPRE01, EspositoGaspPRB05, EspositoGaspJSP05, GarrahanPRL16, GarrahanPRE16}.} 

%now we say what we do:
\comment{In this paper we propose and analyze in great detail a thermodynamically consistent version of the interacting three-state oscillators model. This model can be seen as a toy model for interacting molecular motors \cite{JoannyPRL06}, enzymes \cite{HillPNAS77, HillPNAS81} or switches \cite{RaymoJOC03, SimaoJACS11}.}
At the mean-field level, it displays as a function of the inverse temperature three phases separated by two nonequilibrium phase transitions: a Hopf bifurcation separating a single stable fixed point phase from a stable limit cycle one, and an infinite-period bifurcation separating the limit-cycle phase from a phase with three stable fixed points. At equilibrium only one phase transition survives which separates a phase with a single stable fixed point from one with multiple stable fixed points via a saddle-node bifurcation.
%Should be said but not for here: The first and the latter phase are required for thermodynamic consistency and follow from the high- and low-temperature limit, where the system is entropy- and energy-driven, respectively, and are thus present in both equilibrium and nonequilibrium.
A central result is that the spectrum of the Markovian dynamics generator is shown to encode the information about the two bifurcations that are observed in the mean field. The mean-field dynamics is demonstrated to be characterized by the three eigenvalues with dominant real parts (the null one and a complex conjugated pair). At the Hopf bifurcation, a real-part gap between these eigenvalues and the remaining eigenvalues opens up, enabling the emergence of a metastable mean-field-like oscillatory dynamics over long times. As the second bifurcation is approached, this difference in real parts further increases while the imaginary parts of the dominant eigenvalues significantly drop causing the oscillations to vanish into three metastable fixed points. The fact that the real part of the most dominant complex conjugated eigenvalue pair converges to zero while the gap with respect to the real parts of all other nonzero eigenvalues becomes larger with increasing system size explains the emergence of the mean-field solutions as the perpetuation of the metastable states. 
\comment{After demonstrating the consistency of stochastic thermodynamics across scales (from the microscopic manybody level to the mean field one), we analyze the dissipated work across the different dynamical regimes.} We find that as a function of increasing inverse temperatures the transition towards synchronization is of first order while the outward transition is of second order. A crucial observation is that in the thermodynamic limit, interactions can significantly decrease the dissipated work per oscillator beyond the synchronization threshold and even more so after the second transition towards multistability. Furthermore, interactions in finite assemblies of oscillators enhance this effect in the former case but reduce it in the latter, in particular when the number of oscillators is too low to sustain a long-lasting metastable solution.          
Finally, we demonstrate that when operating as an energy converter, synchronization significantly enhances the power output per oscillator. Despite operating far-from-equilibrium, the efficiency at maximum power remains quite close to the linear-response prediction of $1/2$. 
\comment{Overall, our thermodynamically consistent minimal model for synchronization enables us to reveal with unprecedented detail two complementary facets of a nonequilibrium phase transition: The emergence of different dynamical phases from stochastic dynamics far-from-equilibrium and their thermodynamic characterization using stochastic thermodynamics.} 

The plan of the paper is as follows. First, in Sec. \ref{sec:model}, we introduce the description of our model and perform an exact coarse-graining of the dynamics. Next, Sec. \ref{sec:meanfielddynamics} analyzes the different regimes of the mean-field dynamics which motivate the spectral analysis in \ref{sec:spectralanalysis}. In Sec. \ref{sec:simulations} we compare dynamics between the mean field with finite systems using dynamical Monte Carlo simulations. Furthermore, the thermodynamic laws are formulated in Sec. \ref{sec:thermodynamiclaws} and the work dissipated by noninteracting, small and large interacting networks is compared in \ref{sec:dissipatedwork}. Finally, the power-efficiency trade-off in the mean field is investigated in \ref{sec:efficiencyatmaxpower}. We conclude with a summary and an outlook to proceeding projects in Sec. \ref{sec:conclusion}.

\section{Model}
\label{sec:model}
\subsection{Setup}

\begin{figure}[h!]
  \centering

\includegraphics[width=0.45\textwidth]{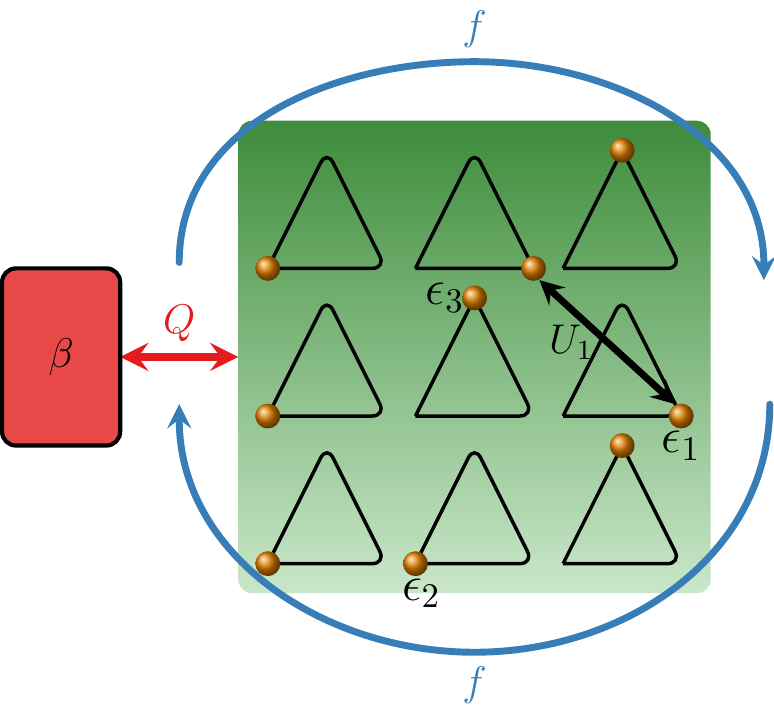}
\caption{Small network of identical and all-to-all interacting three-state units in contact with a heat bath $\invtemperature$ and in the presence of a nonconservative force $\force$. \label{fig:threestatemodelnetwork} }
\end{figure}

We consider a system consisting of $\dimension$ three-state units with energies $\stateenergy_i$ ($i=1,2,3$). Under the constraint of occupying the same state, units interact globally via an interaction potential. The system is subjected to a non-conservative rotational forcing $\force$ and is furthermore in contact with a heat bath at inverse temperature $\invtemperature= (\boltzmann T)^{-1}$, where we set $\boltzmann \equiv 1$ in the following.
The schematics of the setup are depicted in Fig. \ref{fig:threestatemodelnetwork}.

We denote a microstate by the multiindex $\microstate $=$ (\microstate_1,\ldots,\microstate_i,\ldots,\microstate_{\dimension})$ with $\microstate_i \!=\! 1,2,3$. As an example, ordering the units from left to right and from top to bottom, the microstate displayed in Fig. \ref{fig:threestatemodelnetwork} reads $\microstate $=$ (2,1,3,2,3,1,2,2,3)$.
Consider a transition from microstate $\microstate'$ to $\microstate$ amounting to a transition between the single unit energy states $ \stateenergy_j$ to $\stateenergy_i$. For such a transition the occupation numbers change according to $\occupation_i \! \to \! \occupation_i + 1$ and $\occupation_j \! \to \! \occupation_j - 1$.
To determine the change in internal energy $\change \macroenergy(\microstate,\microstate') \!=\! \stateenergy(\microstate) \!- \stateenergy(\microstate') \!+ \Potential(\microstate) \!- \Potential(\microstate')$, the total interaction energy $\Potential(\microstate)$ of the network is required.
Owing to the all-to-all interaction, the total interaction energy is obtained by considering a unit in state $k$ and summing up the remaining number of units occupying the same state. It thus holds
\begin{align}
 \Potential(\microstate) = \frac{u}{N} \sum\limits_{k=1}^3 \; \sum\limits_{l=1}^{\occupation_k\!(\microstate)\!-\!1} \! l 
 = \frac{\potential}{2N} \sum\limits_{k=1}^3 \occupation_k^2(\microstate) + \Potential_0, \label{eq:totalinteraction}
\end{align}
where $\potential/\dimension$ is the interaction strength, $ \Potential_0 \!=\! -\potential \dimension/2 $ is a constant $(\Delta \Potential_0 \!=\! 0)$ and the notation $\occupation_k(\microstate)$ refers to the number of units occupying the single-unit state $k$ in the microstate $\microstate$.
We thus obtain for the change in internal energy
\begin{subequations}
\begin{align}
	\change \macroenergy (\microstate,\microstate')
	\! &= \! \stateenergy(\microstate) \! - \! \stateenergy(\microstate') \! + \! \frac{\potential}{2 \dimension} \! 
	\sum\limits_{k=1}^3 \! \left[ \occupation_k^2(\microstate) \!-\! \occupation_k^2(\microstate') \right] \\
	&= \stateenergy_{i}-\stateenergy_{j} + \frac{u}{\dimension} (\occupation_i - \occupation_j + 1) . \label{eq:changeinternalenergy}
\end{align}
\end{subequations}

\subsection{Master-Equation}

The dissipative dynamics of the system, with the above energetics, is described via a Markovian master equation (ME)
\begin{align} \label{eq:fullmasterequation}
  \dot{\microprobability}_{\microstate}= \sum\limits_{ \microstate' } \microrates_{\microstate\microstate'} \, \microprobability_{\microstate'} \, ,
\end{align}
where $\microprobability_{\microstate}$ denotes the probability to be in the microstate $\microstate$.
The microscopic transition rates $\microrates_{\microstate\microstate'}$ give the probability per unit time for the system to undergo a transition $\microstate'$ to $\microstate$.
With only one transition at a time, it follows that the transition rate matrix is irreducible and stochastic, $ \sum_{\alpha} \microrates_{\microstate\microstate'} = 0$. This implies the existence of a unique stationary state %with probability $\microsteadyprobability_{\microstate}$
\cite{vankampen2007}. We take the microrates to be of Arrhenius form, that is
\begin{align} 
\microrates_{\microstate\microstate'} &= \arrheniusprefactor \; \euler^{-\frac{\invtemperature}{2} \left( \change E(\microstate,\microstate') - \microsign \force \right)}  \, , \label{eq:fulltransitionrates}
\end{align}
with $\Gamma$ setting the timescale. The sign function $\microsign$ gives preference to transitions down the bias $f$ over counteracting ones. It is defined as $\microsign\!=\!1$ for $ \sum_i ( \microstate_i \! - \! \microstate'_i ) \text{ mod } 3 = \! 1$ and $\microsign \! = \! - 1$ otherwise. Furthermore, we emphasize that the rates satisfy local detailed balance
\begin{align} \label{eq:microlocaldetailedbalance}
\ln \frac{\microrates_{\microstate\microstate'}}{\microrates_{\microstate'\microstate}} = - \invtemperature \left( \change E(\microstate,\microstate') - \microsign \force \right)  \, ,
\end{align}
ensuring that the dynamics is thermodynamically consistent. In the long-time limit $t \to \infty$, the system will tend to its unique steady state, $\microsteadyprobability_{\microstate}$, which is in nonequilibrium due to the presence of the non-conservative driving $\force$.

In absence of driving, microscopic detailed balance
\begin{align}
\microrates_{\microstate\microstate'} \, \microeqprobability_{\microstate'} = \microrates_{\microstate'\microstate} \, \microeqprobability_{\microstate} \, ,
\end{align}
holds and along with the local detailed balance condition in Eq. (\ref{eq:microlocaldetailedbalance}) ensures that the equilibrium probability distribution is Gibbsian, \idest,
\begin{align}
\microeqprobability_{\microstate} &= \euler^{-\invtemperature \left( \macroenergy_{\microstate}
- \microfreeenergy^{eq} \right) }  \, ,
% \, .
\intertext{with the equilibrium free energy}
\microfreeenergy^{eq} &= - \frac{1}{\invtemperature} \, \ln \, \sum\limits_{\microstate} 
\euler^{-\invtemperature \macroenergy_{\microstate} } \, .
\end{align}

Formulating the stochastic process as above gives rise to an exceedingly large state space growing exponentially with the number of units in the network as $3^{\dimension}-1$. Yet, a closer inspection reveals that a coarse-graining to a mesoscopic space can be done in which the stochastic dynamics can be represented accurately. In fact, the internal energy (and hence also the microscopic transition rates) $ E(\microstate) \! \equiv \! E(\macrostate)$ does not depend on the topological details encoded in $\microstate$ but only on the mesostate $\macrostate \! \equiv \! (\occupation_1,\occupation_2)$. The number of microstates $\microstate$ belonging to the same mesostate $\macrostate$ is given by
\begin{align} \label{eq:multiplicityfactortrinomial}
\multiplicity(\macrostate) \equiv \sum\limits_{\microstate \in \macrostate} 1 = \binom{\dimension}{\occupation_1} \binom{\dimension - \occupation_1}{\occupation_2} = \frac{\dimension !}{\prod\limits_{i} \occupation_i!}  \,  ,
\end{align}
if the network is made up of $\dimension$ units. Introducing the marginalized probability $\macroprobability_{\macrostate}$ to observe the mesostate $\macrostate$ 
\begin{align} \label{eq:coarsegrainingidea}
\macroprobability_{\macrostate} &\equiv  \sum\limits_{\microstate \in \macrostate } \, \microprobability_{\microstate} \, ,
\end{align}
the ME (\ref{eq:fullmasterequation}) for the full microstate dynamics can be coarse-grained as
\begin{subequations}
\begin{align}
\dot{\macroprobability}_{\macrostate} &= \sum\limits_{\microstate \in \macrostate} \sum\limits_{\macrostate'} \sum\limits_{\microstate' \in \macrostate'} \microrates_{\microstate\microstate'} \, \microprobability_{\microstate'} \\
&= \sum\limits_{\macrostate'} \marginalizedrates_{\macrostate \macrostate'}  \sum\limits_{\microstate \in \macrostate} \sum\limits_{ \microstate' \in \macrostate' } \; \microprobability_{\microstate'} \, \chi_{\microstate',\microstate} \label{eq:coarsegrainingcrucialline} \\
&= \sum\limits_{\macrostate'} \rates_{\macrostate \macrostate'}\; \macroprobability_{\macrostate'} \, , \label{eq:macromasterequation}
\end{align}
\end{subequations}
with the marginalized mesoscopic transition rates $\rates_{\macrostate \macrostate'} = \multiplicity(\macrostate,\macrostate') \marginalizedrates_{\macrostate \macrostate'}$.
We note that the coarse-graining preserves the stochastic property and the irreducibility of the transition rate matrix.

The characteristic function $\chi_{\microstate',\microstate}$ emerging in Eq. (\ref{eq:coarsegrainingcrucialline}) is a result of pulling the sums through the microscopic transition rate matrix since the information that $\microrates_{\microstate  \microstate'} \neq 0$, only if transitions between $\microstate'$ and $\microstate$ are possible, would be lost.
Consequently, the function takes the value 1 if $\microstate'$ and $\microstate$ are connected, and is 0 otherwise.

To determine the constrained multiplicity factor $\multiplicity(\macrostate,\macrostate') \equiv 
\sum_{\microstate \in \macrostate} \, \chi_{\microstate',\microstate}$, we need to address the question of how many microstates $\microstate \! \in \! \macrostate$ are connected with $\microstate' \! \in \! \macrostate'$.
Two macrostates are connected, if two of three occupation numbers $(\occupation_1,\occupation_2,N\!-\!\occupation_1\!-\!\occupation_2)$ of the macrostates differ by exactly one. In the microstate space, this corresponds to, if, compared entrywise, exactly one number being different in the tuples representing the two microstates. Thus we obtain for the constrained multiplicity factor
\begin{align}
\hspace*{-.1cm} \multiplicity(\macrostate,\macrostate')
 &= \occupation'_1\, \delta_{\occupation'_1,\occupation_1+1} + \occupation'_2 \, \delta_{\occupation'_2,\occupation_2+1} + \nonumber \\
 &+ \! (\dimension \!\!-\!\! \occupation'_1 \!\!-\!\! \occupation'_2) \, \delta_{\dimension \!- (\occupation'_1 + \occupation'_2), \dimension \!- (\occupation_1 \!+\! \occupation_2)+1} \, ,
\end{align}
which indeed does not require any microscopic information. Hence the coarse-graining of the dynamics is exact and leads to a closed ME (\ref{eq:macromasterequation}) represented in terms of mesoscopic states $\macrostate$.
This coarse-graining significantly reduces the dimensionality of the state space which grows as $ \left[ (\dimension+1)(\dimension+2) \right] /2 -1 $, thus quadratically as $\dimension$ becomes large. Using Boltzmann's entropy
\begin{align} \label{eq:internalentropy}
\internalentropy(\macrostate) = \ln \multiplicity(\macrostate)\, ,
\end{align}
the multiplicity factor of the microstates can be incorporated into the macrorates in a physically meaningful way.
The mesoscopic local detailed balance relation thus reads
\begin{align} \label{eq:macrolocaldetailedbalance}
\ln \frac{ \rates_{\macrostate\macrostate'} }{  \rates_{\macrostate'\macrostate} } = - \invtemperature \left[\change \macrofreeenergy(\macrostate,\macrostate')  - \macrosign \force \right] ,
\end{align}
with $\change \macrofreeenergy(\macrostate,\macrostate') \!=\! \Delta \macroenergy(\macrostate,\macrostate') \! - \! \invtemperature^{-1} \change \internalentropy(\macrostate,\macrostate') $ being the difference in free energy  between the macrostates $\macrostate'$ and $\macrostate$. \newcomment{The mesoscopic sign function $\macrosign$ is defined analogously to $\microsign$. Thus, $\macrosign=1$ for $ \sum_i ( \macrostate_i \! - \macrostate_i' ) \text{ mod } 3 = \! 1$ and $\macrosign \! = \! - 1$ otherwise.}

The mesoscopic local detailed balance relation (\ref{eq:macrolocaldetailedbalance}) implies that at $t \to \infty$ and for $\force=0$ the mesoscopic detailed balance
\begin{align}
\macrorates_{\macrostate\macrostate'} \, \macroeqprobability_{\macrostate'} = \macrorates_{\macrostate'\macrostate} \, \macroeqprobability_{\macrostate} \, ,
\end{align}
holds and the mesoscopic equilibrium probability distribution
\begin{align} \label{eq:mesoscopicequilibriumdistribution}
\macroeqprobability_{\macrostate} = \euler^{-\invtemperature \left(\macrofreeenergy_{\macrostate} - \macrofreeenergy^{eq} \right) } ,
\end{align}
is again of the Gibbs form with the equilibrium free energy
\begin{align}
\macrofreeenergy^{eq} = -\frac{1}{\invtemperature} \ln \sum\limits_{\macrostate} \euler^{-\invtemperature \macrofreeenergy_{\macrostate} } \, .
\end{align}

\section{Mean-Field Dynamics}
\label{sec:meanfielddynamics}

In order to further reduce the complexity of the state space of the mesoscopic ME (\ref{eq:macromasterequation}) we first operate in the mean-field (MF) limit where $\dimension \!\to\! \infty$. In this limit, the total change in internal energy due to a transition in Eq. (\ref{eq:changeinternalenergy}) simplifies and the corresponding scaled current density  $J(\occupationdensity_i,\occupationdensity_j) \! \equiv \! \lim_{\dimension \to \infty} \macrorates_{\macrostate \macrostate'} / \dimension $ becomes
\begin{align}
J(\occupationdensity_i,\occupationdensity_j) = \arrheniusprefactor \, \euler^{-\frac{\invtemperature}{2}\! \left( \stateenergy_i - \stateenergy_j + \potential(\occupationdensity_i - \occupationdensity_j) - \meanfieldsign \force \right) } \, \occupationdensity_j \, ,
\end{align}
where $\occupationdensity_i \!=\! \occupation_i/\dimension$ denotes the occupation density of the single-unit state $i$
\newcomment{and $\meanfieldsign=1$ for $ (i-j) \text{ mod } 3 = \! 1$, while $\meanfieldsign \! = \! - 1$ otherwise.}
 The evolution equation for the mean occupation density reads
\begin{align}
\langle \dot{\occupationdensity}_i \rangle &= \sum\limits_{j \neq i} \langle J(\occupationdensity_i,\occupationdensity_j) \rangle - \langle J(\occupationdensity_j,\occupationdensity_i) \rangle .
\end{align}
In the MF approximation we replace any $n$-point correlation function with a product of $n$ averages thus yielding
\begin{align}
\dot{ \meanfieldprobability}_i \equiv \langle \dot{\occupationdensity}_i \rangle &= \sum\limits_{j \neq i} J(\langle \occupationdensity_i \rangle ,\langle \occupationdensity_j \rangle ) - J(\langle \occupationdensity_j \rangle ,\langle \occupationdensity_i \rangle ) , 
\label{eq:meanfieldapproximation}
\end{align}
which represents a closed nonlinear equation. The validity of this approximation can be proved \cite{vankampen2007} in the macroscopic limit $\dimension \! ~ \to ~ \! \infty$. Hence the MF system can be described by a single three-state unit, where the (average) occupation density of the single-unit states is assigned to the three states of the MF unit. \comment{We therefore identify the MF occupation density, $ \meanfieldprobability_i$, as the probability for any unit to occupy the single-unit state $i=1,2,3$. Its dynamics is ruled by the nonlinear MF equation}
\begin{align}
\dot{ \meanfieldprobability }_i \! &= \sum\limits_{j } \meanfieldrates_{ij} \, \meanfieldprobability_j \, , \label{eq:meanfieldmasterequation}
\end{align}
\newcomment{with the MF transition rates 
\begin{align}
\meanfieldrates_{ij} = \arrheniusprefactor \, \euler^{-\frac{\invtemperature}{2}\! \left( \stateenergy_i - \stateenergy_j + \potential(\meanfieldprobability_i - \meanfieldprobability_j) - \meanfieldsign \force \right) } \, ,
\end{align}}
obeying local detailed balance
\begin{align} \label{eq:meanfieldlocaldetailedbalance}
\ln \frac{ \meanfieldrates_{ij}}{ \meanfieldrates_{ij}} &= -\invtemperature \left( \stateenergy_i-\stateenergy_j + \potential(\meanfieldprobability_i-\meanfieldprobability_j)- \meanfieldsign \, \force \right) .
\end{align}
\comment{Unit conservation erases one degree of freedom such that there are only two independent variables $\occupationdensity_1$ and $\occupationdensity_2$.}
We proceed by choosing a flat energy landscape, \idest by setting $\stateenergy_i \!=\! \const \, \forall i$. This allows us to immediately read off the symmetric point $\meanfieldprobability_i^* = 1/3$ as an analytic solution to the nonlinear MF Eq. (\ref{eq:meanfieldmasterequation}). Linearizing the Eq. around this fixed point (FP), $ \dot{\meanfieldprobability}_i = \sum\limits_{j } \left. \tfrac{\D \meanfieldrates_{ij}}{\D \meanfieldprobability_j} \right|_{\meanfieldprobability_j = \meanfieldprobability_j^*} \meanfieldprobability_j$, we find for the eigenvalues of the linear stability matrix 
\begin{align}
  \meanfieldeigenvalue_{\pm} \!=\! -\arrheniusprefactor ( \invtemperature \potential \!+\! 3)
  \cosh \left(\frac{\invtemperature \force}{2}\right) \! \pm \! \i \sqrt{3} \,  \arrheniusprefactor \sinh \left(\frac{\invtemperature \force}{2}\right) . \label{eq:linearstabilityanalysis}
\end{align}
For attractive interactions ($\potential<0$) between the units the real part of $\meanfieldeigenvalue_{\pm}$ changes its sign at $\invtemperature_{c_1} = -3/u$. This crossover suggests that the stable symmetric FP destabilizes and degenerates into a limit cycle (LC) corresponding to a Hopf bifurcation indicative of synchronization.

%Numerically, we find a critical forcing, $\force_c \approx 0.21$, below which this bifurcation does not appear because the imaginary part in Eq. (\ref{eq:linearstabilityanalysis}) is shrinking such that there can not be stable oscillations.

In appendix \ref{sec:hopfbifurcationproof}, the LC is characterized in the vicinity of the Hopf bifurcation which is shown to occur supercritical for attractive interactions. Moreover, a closer inspection of the MF rates in Eq. \eqref{eq:meanfieldlocaldetailedbalance} reveals that for any $\force$ and $\invtemperature$ repulsive interactions, $\potential >0$, always lead to the stable symmetric FP.

\begin{figure}[h!]
\begin{center}
\includegraphics[width=0.45\textwidth]{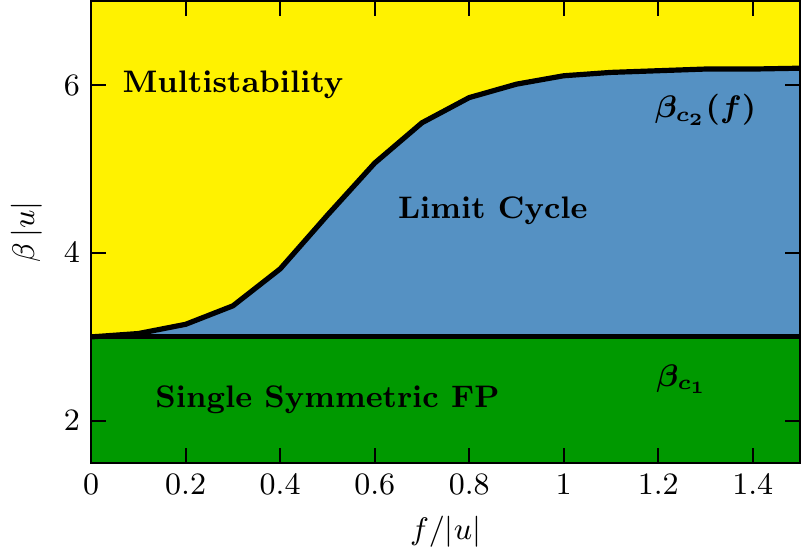}
\caption{\comment{Phase space in the MF varying the parameters $\invtemperature$ and $\force$ in units of $\potential$. The black lines correspond to the set of critical points $\invtemperature_{c_1}$ and $\invtemperature_{c_2}(\force)$.} \label{fig:meanfieldparameterphasespace} }
\end{center}
\end{figure}

\comment{Fig. \ref{fig:meanfieldparameterphasespace} depicts the MF phase space for different $\invtemperature$ and $\force$ in units of $\potential$. The symmetric fixed point is only stable for $\invtemperature < \invtemperature_{c_1}$. We find in agreement with Eq. \eqref{eq:linearstabilityanalysis} that for finite $\force$ there is a phase characterized by stable LCs if $ \invtemperature \geq \invtemperature_{c_1}$. For any value of $\force$, there is an additional phase with three non-symmetric FPs for $\invtemperature \geq \invtemperature_{c_2}(\force)$. }

\noindent

\comment{We set $\potential = -1$ in the following and briefly address a subtlety of the MF system.
In Fig. \ref{fig:meanfieldparameterphasespace} the analytic solution to Eq. \eqref{eq:meanfieldmasterequation}, $\meanfieldprobability_i=1/3$, is chosen as initial condition.
In fact, at temperatures \newcomment{close to the} first critical temperature $\invtemperature_{c_1}$ the long-time solution is initial-condition dependent: For $0 < \force < \force_c \approx 0.21 $ there is a finite set of initial conditions different from the symmetric FP that will \emph{not} lead to a LC but to a non-symmetric stable fixed point. If $\force \geq \force_c$, the dynamics will always exhibit a LC regardless of the chosen initial condition. }

\begin{figure}[h!]
\def\xlabel{2.6}
\def\ylabel{3.5}
\begin{center}

\includegraphics[width=0.215\textwidth]{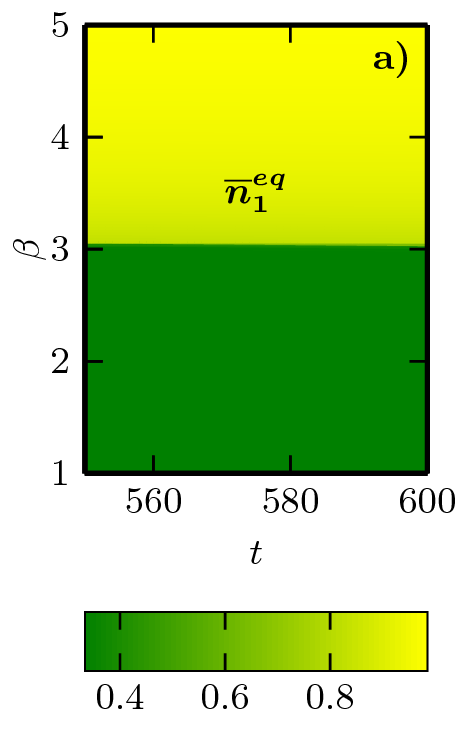}
\includegraphics[width=0.255\textwidth]{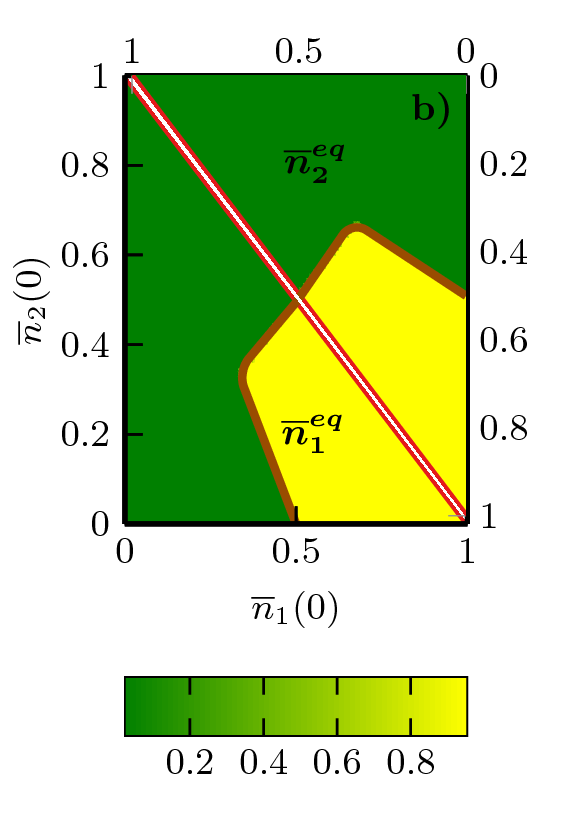}
\caption{Density plot of the equilibrium occupation density $\meanfieldeqprobability_1$ for different $\invtemperature$ and times $t$ \comment{for an initial condition $\bm{\meanfieldprobability}(0)$ equal to the symmetric FP in a)}, and as a function of all physical initial conditions $\meanfieldprobability_1(0)$ and $\meanfieldprobability_2(0)$ at time $t=10^3$ and for $\invtemperature = 4.0$ in b) (lower left triangle). For completeness, the upper right triangle in panel b) shows the other component $\meanfieldeqprobability_2$. The times are chosen to be sufficiently long such that the system has relaxed to equilibrium. \label{fig:equilibriummeanfieldphasespace} }
\end{center}
\end{figure}

Before studying the different nonequilibrium phases of this model, we discuss it for $\force \!=\! 0$, \idest at equilibrium. Figure \ref{fig:equilibriummeanfieldphasespace}a) shows\comment{, starting from the initial condition $\bm{\meanfieldprobability}(0) = (1/3,1/3)^{\top}$,} the long-time solution $\meanfieldeqprobability_1(t)$ for different values of $\invtemperature$.

\noindent

\comment{At the critical temperature $ \invtemperature_{c_1} $ the system exhibits three non-symmetric stable FPs that emerge via a saddle-node bifurcation.} Our thermodynamic framework allows us also to work within the nomenclature of statistical mechanics. Interestingly, the saddle-node bifurcation corresponds to a first-order equilibrium phase transition since the derivative of the MF free energy with respect to $\invtemperature$ at the critical point $\invtemperature_{c_1}$ is divergent. Starting from the symmetric FP, these attractive FPs are observed to move towards the corners of the triangle in the $\meanfieldeqprobability_1\!-\!\meanfieldeqprobability_2$ plane. This is physically plausible since at low temperatures the system tends to occupy its lowest energy state where all units are occupying the same state. The dependence of the multiple equilibrium states on the initial condition in the low-temperature phase is investigated in Fig. \ref{fig:equilibriummeanfieldphasespace}b).

In the lower triangle, $\meanfieldeqprobability_1$ is plotted as a function of all physical initial conditions ($\meanfieldprobability_1(0)$,$\meanfieldprobability_2(0)$). As a complement, the other MF probability $\meanfieldeqprobability_2$ is shown in the upper triangle, where the axis labels are omitted for better readability. Each triangle exhibits two phases which are separated by a contour line. Combining these two panels \footnote{Note that a state does not correspond to a folding of the two triangles but a rotation of one of the two planes about $180^{\circ}$ and subsequent overlapping of the two layers.}, we find that for every physical initial condition the system will eventually arrive at one of the three non-symmetric FPs. These differ only by permutations of their components $(\meanfieldeqprobability_1,\meanfieldeqprobability_2,1-\meanfieldeqprobability_1-\meanfieldeqprobability_2)$, where two of them are identical according to the two phases in each of the panels in b).

%As already mentioned, there is a critical forcing, $\force_c \approx 0.21$, below which there are no major qualitative changes between nonequilibrium and equilibrium. If the forcing exceeds this threshold value a third phase characterized by stable oscillations emerges for intermediate temperatures. Hence for the Hopf bifurcation to appear the system needs to be driven far-out-of equilibrium.

\begin{figure}[h!]
\def\xlabel{2.6}
\def\ylabel{3.5}
\begin{center}

\includegraphics[width=0.195\textwidth]{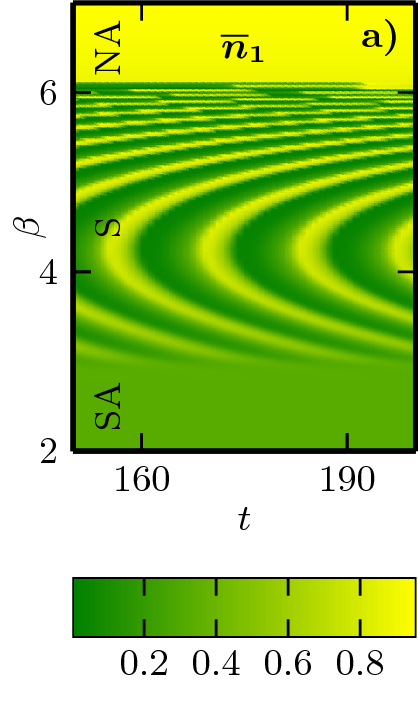}
\includegraphics[width=0.26\textwidth]{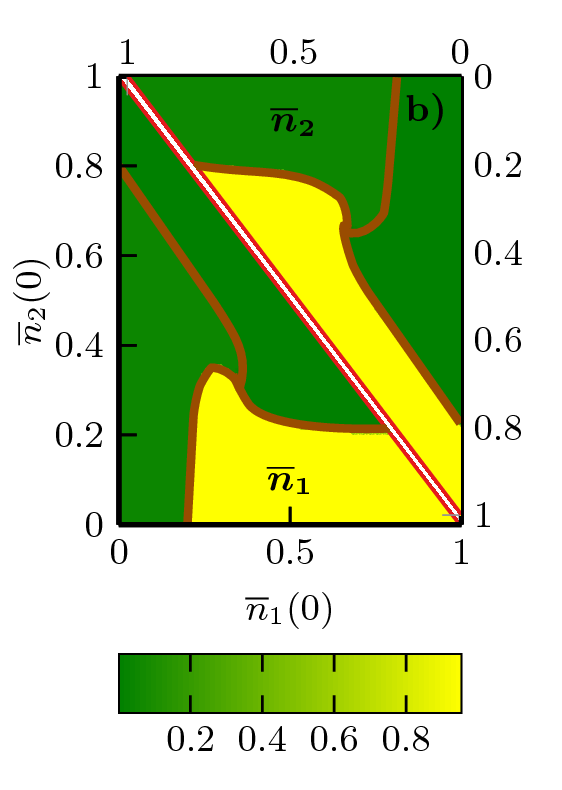}
\caption{ Illustration of the occupation probability $\meanfieldprobability_1$ as a function of $\invtemperature$ and $t$ for $f\!=\!1.0$ in a), while b) shows the occupation densities $\meanfieldprobability_1$ in the lower left triangle and $\meanfieldprobability_2$ in the upper right triangle as a function of all initial conditions at $\meanfieldprobability_1(0)$ and $\meanfieldprobability_2(0)$ at time $t \!=\! 10^3$ and for $\invtemperature \!=\! 7.0$. The initial condition underlying the plots in panels a) is $ (\meanfieldprobability_{1}(0)=1,\meanfieldprobability_2(0)=0) $. The times are chosen such that the system has reached either the unique FP in the symmetric asynchronous phase (SA), the stable LC in the synchronous phase (S), or one of the three non-symmetric FPs in the non-symmetric asynchronous phase (NA). The triple points defined by the intersecting contour lines in b) correspond to the symmetric unstable FP present in the NA phase. \label{fig:meanfieldphasespace} }
\end{center}
\end{figure}

\comment{Figure \ref{fig:meanfieldphasespace}a) depicts the MF probability $\meanfieldprobability_1(t)$ as a function of $\invtemperature$ at long times for $\force=1.0$ at which the range of $\invtemperature$ values for which LCs can be observed is close to its maximum, according to Fig. \ref{fig:meanfieldparameterphasespace}. In agreement with Eq. (\ref{eq:linearstabilityanalysis}), the oscillations emerge at $\invtemperature_{c_1}$ for any finite $\force$.} The oscillations exhibit an increasing frequency with $\invtemperature$ up to a point where they slow down. At the second critical point, $\invtemperature_{c_2}(\force = 1.0) \approx 6.11$, the oscillation period diverges corresponding to an infinite-period bifurcation \cite{keener1981siam}. The initial-condition dependence of the stationary states in the non-symmetric asynchronous phase (NA) is depicted in d), with $ \beta = 7.0$. Again, depending on the chosen initial condition, the system will eventually arrive in one of the three non-symmetric FPs, which are again related to each other by permutations of their components. Here, in contrast to the equilibrium case, all components are different. This reflects the presence of the force distorting the symmetry of the states. The distortion occurs since it is more likely to jump from the largely populate state into the lower occupied state following the bias rather than the opposite way. This asymmetry naturally increases as the system is driven further out-of-equilibrium.

This constitutes the first important result. We have developed a minimal model which, according to Eqs. (\ref{eq:linearstabilityanalysis}) and (\ref{eq:lyapunovresult}), exhibits synchronization and is thermodynamically consistent due to Eqs. (\ref{eq:microlocaldetailedbalance}), (\ref{eq:macrolocaldetailedbalance}) and (\ref{eq:meanfieldlocaldetailedbalance}). We also note that synchronization only occurs in a finite range of temperatures: Fig. \ref{fig:equilibriummeanfieldphasespace} illustrates that at low temperatures the equilibrated system is energy-driven and tends to its energetic ground state, while for very high temperatures the system is entropy-driven and takes a uniform stationary probability distribution. By extrapolation from equilibrium to the non-equilibrium scenario where the synchronization phase emerges, we realize that Fig. \ref{fig:meanfieldphasespace} invites for an analogous physical interpretation of the low- and high-temperature limit in the non-equilibrium case. Moreover, the limit $\invtemperature \to 0$ represents equilibrium since forward and backward transition for each pair of states becomes equally probable for any $\force$ and thus detailed balance holds. We remark furthermore that the term ``minimal'' refers to the dimensionality of the MF dynamics given by Eq. (\ref{eq:meanfieldmasterequation}), which is a natural requirement to observe synchronization since a single-variable nonlinear differential equation can naturally not have complex eigenvalues.

\section{Spectral Analysis: Metastability}
\label{sec:spectralanalysis}

A crucial aspect of our model is that it allows us to study its (thermo-)dynamic features for large but finite system sizes and in particular to monitor the convergence of the stochastic dynamics to the MF dynamics. In order to proceed, we begin by stating the formal solution to the mesoscopic ME (\ref{eq:macromasterequation}) that reads
\begin{subequations}
\begin{align}
\macroprobabilityvector(t) &= \euler^{\macroratematrix t} \, \macroprobabilityvector(0) \\
&= \! \sum\limits_{i,i^*} \! \euler^{\masterequationeigenvalue_i t} \! \underbrace{ \lefteigenvectors_{i}  \macroprobabilityvector(0) }_{\equiv c_i} \righteigenvectors_i + c_{i^*}
\euler^{\masterequationeigenvalue_{i^*} t} \righteigenvectors_{i^*}  ,
\label{eq:spectraldecomposition}
\end{align}
\end{subequations}
where $\macroprobabilityvector(0)$ is the initial probability distribution, $\masterequationeigenvalue_i$ are the eigenvalues and $ \lefteigenvectors_i $, $\righteigenvectors_i$ are the left- and right eigenvectors of the non-symmetric real transition rate matrix $\macroratematrix$ constituting an orthonormal dual basis $\lefteigenvectors_i \righteigenvectors_j = \delta_{ij} $. The index $i^*$ characterizes, if existent, the modes with eigenvalues and eigenvectors being the complex-conjugated to those labeled with $i$.
The Perron-Frobenius theorem (PFT) \cite{vankampen2007} stipulates that for this irreducible, autonomous and stochastic matrix there is a non-degenerate eigenvalue, the Perron-Frobenius eigenvalue (PFE), $\masterequationeigenvalue_0=0$, which is strictly greater than the real part of any other eigenvalue, $ \Real( \masterequationeigenvalue_i ) < \masterequationeigenvalue_0  \; \forall i\neq 0$. Note that the labeling of the eigenvalues is given by the order of their real parts,
$ 0 > \Real(\masterequationeigenvalue_{1}) > \ldots \;$.
%The PFE has an associated right eigenvector $\left| \Phi_0^R \right\rangle$ and left eigenvector $\left\langle \Phi_0^L \right|$ which are both real and the only positive eigenvectors with a finite norm. %Moreover, the left eigenvector associated with the PFE is flat and thus acts as a normalization factor.
Consequently, Eq. (\ref{eq:spectraldecomposition}) has a unique, infinite-time solution, $\macroprobabilityvector^s= c_0 \righteigenvectors_0 $, characterized by the PFE and the associated right eigenvector $\Phi_0^R$.

Hence the stationary state of the mesoscopic system $\macroprobabilityvector^{s}$ cannot exhibit stable oscillations (S phase) or multistability (NA phase). On the other hand, one expects that the transition from the mesoscopic system to the MF is smooth as the system size $\dimension$ grows. This apparent paradox is caused by the non-commutation of the infinite-time limit $t \to \infty$ and the mean-field limit $\dimension \to \infty$, \idest
\begin{align}
\lim_{t \to \infty}  \lim_{\dimension \to \infty} \; \macroprobabilityvector(t)  \neq  \lim_{\dimension \to \infty} \; \underbrace{ \lim_{t \to \infty} \; \macroprobabilityvector(t) }_{ \macroprobabilityvector^s } , \quad \text{if } \; \invtemperature \geq \invtemperature_{c_1} .
\end{align}
The right-hand side corresponds to the symmetric stationary state of the SA phase for all temperatures, while the left-hand side is temperature dependent: For $\invtemperature_{c_1} \leq \invtemperature < \invtemperature_{c_2}$ the system is in a time-periodic state (S phase) and for $ \invtemperature \geq \invtemperature_{c_2}$ the dynamics will go to one of the non-symmetric steady states (NA phase) depending on the chosen initial condition. At $\invtemperature < \invtemperature_{c_1}$ the left-hand side also corresponds to the symmetric stationary state, hence the two limits commute only at sufficiently high temperatures.

To resolve this apparent contradiction we look for clues in the spectrum of the Markov generator in the mesoscopic ME (\ref{eq:macromasterequation}) and establish a link between finite-size systems and MF via the notion of metastability.
Even though the stationary state is inevitably reached in the infinite-time limit, there could be long-living metastable states that display the phenomenology of the MF. The time-scales to characterize such a state are encoded in the spectrum as follows
\begin{subequations}
\begin{align}
\tau_{r} & \sim - \frac{1}{\Real(\masterequationeigenvalue_{1})} \label{eq:relaxationtime} \\
\tau_{m} & \sim - \frac{1}{\Real(\masterequationeigenvalue_{2})} \label{eq:metastabletime} \\
\tau_{l} & \equiv \tau_{r} - \tau_{m} \sim  \frac{1}{|\Real(\masterequationeigenvalue_{1})|} 
\left( 1 - \frac{\Real(\masterequationeigenvalue_{1})}{\Real(\masterequationeigenvalue_{2})} \right), 
\label{eq:lifetime}
\end{align}
\end{subequations}
where $\tau_{r}$ is the relaxation time to reach the unique steady state, \idest it specifies the time-scale at which all finite-time modes have been removed from the dynamics. $\tau_{m}$ is the metastable time at which all modes have decayed except for those forming the metastable state, that is the one associated with the eigenvalue $\masterequationeigenvalue_{1}$ and the stationary one characterized by the PFE $\masterequationeigenvalue_0$. Here, we assume that only a pair of modes associated with a complex-conjugated non-null eigenvalue is contributing to metastability, while there could be an arbitrary number of modes forming the metastable state. This assumption will be numerically verified in the following.

Physically, $\tau_{l}$ corresponds to the lifetime of that metastable state. To reconcile the stochastic dynamics with its asymptotic solution in the macroscopic limit, the MF dynamics, $\tau_l$ is required to become increasingly larger with the system size $\dimension$, while $\tau_m$ remains finite since the different MF phases emerge at finite time. Using Eqs. \eqref{eq:relaxationtime}--\eqref{eq:lifetime}, these prerequisites translate into conditions on the real parts of the dominant eigenvalues of the Markov generator: The real-part gap between the two first non-null eigenvalues, $\Real(\masterequationeigenvalue_1) - \Real(\masterequationeigenvalue_2)$, has to increase by $\Real(\masterequationeigenvalue_1)$ converging to zero (corresponding to a diverging relaxation time $\tau_r$), while $\Real(\masterequationeigenvalue_2)$ has to approach a finite value (assuring the emergence of the metastable phenomena at finite times). Moreover to mimic MF dynamics the metastable state has to be oscillatory ($\Imag(\masterequationeigenvalue_1)\neq 0$) in the S phase and quasistationary ($\Imag(\masterequationeigenvalue_1) = 0$) in the NA phase.

\begin{figure}[h!]
\begin{center}
\includegraphics[width=0.46\textwidth]{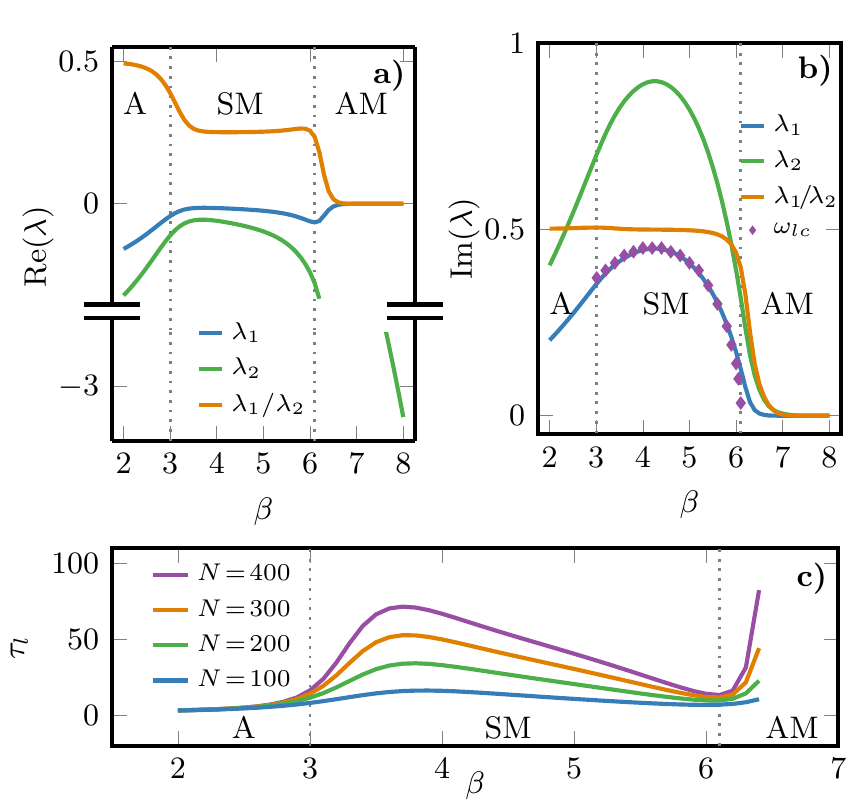}

\caption{The real part a) and the imaginary part b), as well as the ratio, of the two most dominant eigenvalues with distinct real part, $\masterequationeigenvalue_1$,$\masterequationeigenvalue_2$, with positive imaginary part are depicted as a function of $\invtemperature$. \comment{In addition, the LC frequency $ \limitcyclefrequency $ that is numerically extracted from the asymptotic ($t\to \infty$) MF dynamics is compared to the imaginary part of the most dominant eigenvalue.} All eigenvalues correspond to a generator $\W$ for a system of size $\dimension=~300$. Panel c) shows the lifetime of the metastable state $\tau_l$ as function of $\invtemperature$ for different system sizes. The labels of the different phases, that is the asynchronous phase (A), the synchronous metastable phase (SM) and the asynchronous metastable phase (AM) are in correspondence with the labels of the different phases in the MF limit introduced in the preceding Sec. \ref{sec:meanfielddynamics}.
\label{fig:eigenvaluesvsbeta}}
\end{center}
\end{figure}

Before addressing the question of how the stochastic dynamics converges to the MF, we study the real parts a) and the imaginary parts b) of the two dominant non-zero eigenvalues of the spectrum in all three different phases ($2 \leq \invtemperature \leq 8 $) for a system size of $\dimension = 300$ in Fig. \ref{fig:eigenvaluesvsbeta}. We remark that for all $\invtemperature$, these two eigenvalues in fact occur as complex-conjugated pairs and only those with a positive imaginary part are depicted in panel b). Furthermore, to stress that the different phases of the finite-size system for $\invtemperature > \invtemperature_{c_1}$ are only present for finite times, we rename them differently than in the MF: asynchronous phase (A), synchronous metastable phase (SM) and asynchronous metastable phase (AM).

\noindent

As can be seen in panel a), the real parts of the two eigenvalues both approach zero up to $\invtemperature \approx 4$ followed by a monotonic decrease of $\Real(\masterequationeigenvalue_2)$ while $\Real(\masterequationeigenvalue_1)$ changes only slightly and for $\invtemperature > \invtemperature_{c_2}$ rapidly goes to zero. According to Eq. \eqref{eq:lifetime}, this observation along with the fact that $\Real (\masterequationeigenvalue_1) / \Real (\masterequationeigenvalue_2) $ drops at both critical points (dashed lines) suggests that the lifetime $\tau_l$ of the metastable state is increasing from the SM to the AM regime. The small values of $|\Real(\masterequationeigenvalue_{1})|$ in the SM and AM phase and the sharp changes in the ratio of the real parts at both critical points provide a first hint that the metastable state is constituted by only the stationary mode and those associated with the first complex-conjugated non-null eigenvalue.

This claim is further strengthened by studying the corresponding imaginary parts of these eigenvalues as shown in Fig.  \ref{fig:eigenvaluesvsbeta}b). \comment{We find an excellent agreement in the SM phase between the LC frequency $\limitcyclefrequency$ in the MF that is numerically extracted from the dynamics and $\Imag (\masterequationeigenvalue_1 )$. The LC frequency $\limitcyclefrequency$ only coincides with the imaginary part of the Jacobian from the linear stability analysis in Eq. \eqref{eq:linearstabilityanalysis} at the bifurcation point $\invtemperature_{c_1}$, where the linearization of the nonlinear ME \eqref{eq:meanfieldmasterequation} is exact.} Moreover, the ratio between the imaginary parts of $\masterequationeigenvalue_1$ and $\masterequationeigenvalue_2$ remains nearly constant at $0.5$ within the A and SM phase implying that the frequency of oscillation of the mode corresponding to $\masterequationeigenvalue_2$ is half as that of $\masterequationeigenvalue_1$. In the AM phase $\Imag (\masterequationeigenvalue_1 )$ quickly goes to zero consistent with our MF observations that show no oscillations.

Consistent with the discussion of the real parts, Fig. \ref{fig:eigenvaluesvsbeta}c) illustrates that the lifetime of the metastable state is nearly zero in the A phase and starts to increase significantly at the first critical point up to a local maximum in the SM phase. The lifetime $\tau_l$ is monotonically decreasing for larger $\invtemperature$ before it sharply rises in the AM phase.
All clues thus indicate that in the two phases where the MF exhibits non-unique solutions at infinite times, the finite system displays metastability. As expected, for all temperatures in the metastable phases the lifetime is monotonically increasing with $\dimension$.

\begin{figure}[h!]
\begin{center}

\includegraphics[width=0.47\textwidth]{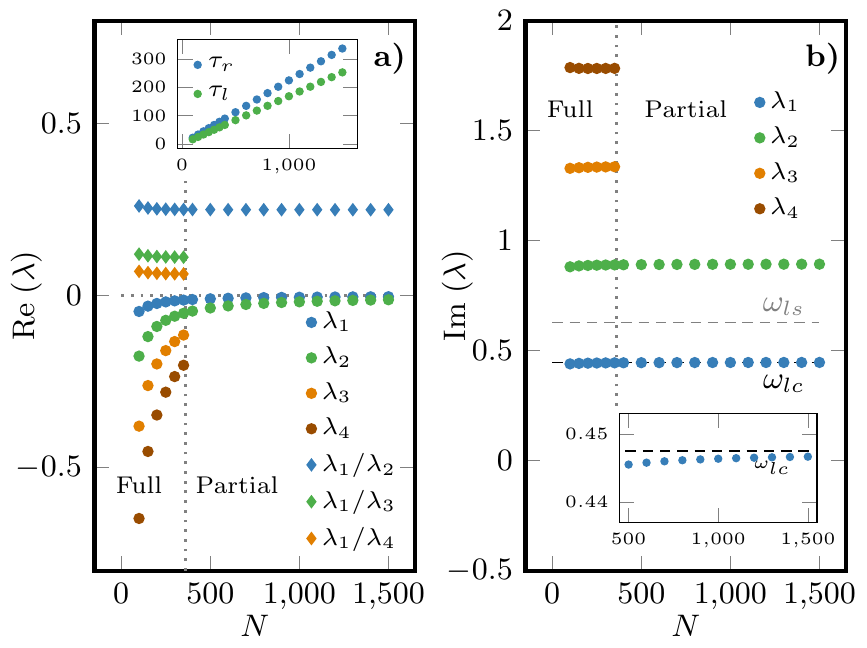}
\includegraphics[width=0.46\textwidth]{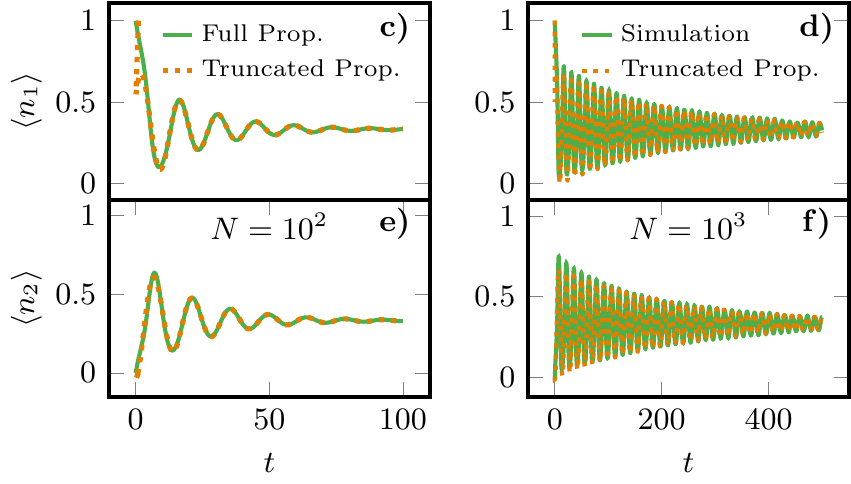}

\caption{Real a) and corresponding imaginary b) parts of the four most dominant eigenvalues with distinct and finite real part for different $\dimension$ and for $\invtemperature \!=\! 4$ as a representative of the SM phase. The data points corresponding to system sizes larger than $\dimension=350$ are not resulting from a full diagonalization of the matrix but were obtained exploiting the sparseness of the matrices (maximal 6 of the approximately $ \! \dimension^2/2$ entries of every row/column are nonzero), using a recursive algorithm, to obtain the dominant part of the spectrum. \comment{The inset depicts the relaxation time scale $\tau_r$ and the lifetime of the metastable state $\tau_l$ as a function of $\dimension$.} In b) the dashed, horizontal lines labeled as $\limitcyclefrequency$,$\linearstabilityfrequency$ correspond to the LC frequency in the MF and to the imaginary part of the linear stability matrix eigenvalue from Eq. (\ref{eq:linearstabilityanalysis}), respectively.
The mean occupation density $ \langle \occupationdensity_i \rangle$ as a function of time for both the full (\ref{eq:spectraldecomposition}) and truncated (\ref{eq:truncatedspectraldecomposition}) propagation [$i$=1 in c), d) and $i$=2 in e) and f)] for the different network sizes $\dimension=10^2,10^3$. The dynamics for $\dimension \!=\! 10^3$ was generated using the direct Gillespie method. \label{fig:dominanteigenvalues} }
\end{center}
\end{figure}

Next, to shed some light on the convergence of the finite-system dynamics to the MF dynamics, we investigate the changes in the spectrum as we approach the MF limit. To this end, we look at the first few dominant non-zero eigenvalues as a function of the system size $\dimension$ at $\invtemperature = 4$ representing the SM phase. We observe in Fig.~\ref{fig:dominanteigenvalues}a) that the real parts of these eigenvalues are approaching the PFE. \comment{Though the inset reveals an increasing time-scale separation between the mode associated with $ \masterequationeigenvalue_1$ and the faster decaying modes for larger systems. The monotonically increasing behavior of $\tau_l$ and $\tau_r$ with $\dimension$ implies an increasing lifetime of the metastable state, while this time window is shifted to increasingly larger times, hence the finite-system dynamics are converging to the MF.}
To be fully consistent with the MF, the metastable state must be appearing in the dynamics at a finite time. Taking into account all the aforementioned hints (encoded in Fig. \ref{fig:eigenvaluesvsbeta} and to be made in the following) that indeed only the modes associated with $\masterequationeigenvalue_{1,1^*}$ are contributing to the metastability and therefore in correspondence with the MF solution, it is reasonable to expect that $ \Real (\masterequationeigenvalue_2) $ converges to a finite value for larger $\dimension$.
Unfortunately, extracting the dominant eigenvalues of the generator for even larger $\dimension$ is not feasible.

\noindent

As another striking evidence for the hypothesis that the metastable state comprises only the stationary and the first non-null complex-conjugated mode, the imaginary part of the dominant eigenvalue $\masterequationeigenvalue_1$ smoothly converges to the LC frequency $\limitcyclefrequency$ in the MF while the imaginary parts of other modes display a distinct separation as seen in Fig.~\ref{fig:dominanteigenvalues}b). This is confirmed in Figs. \ref{fig:dominanteigenvalues}c) -- f) depicting the mean occupation densities, $\langle \bm{\occupationdensity} (t) \rangle ~=~  \sum_{\macrostate}  \macrostate / \dimension \, \macroprobability_{\macrostate}(t)$, using the full spectral decomposition of the Markov generator in Eq. (\ref{eq:spectraldecomposition}) and the truncated one
\begin{align}
\macroprobabilityvector(t) \!\overset{t \gg \tau_m}{\approx}\! c_0   \righteigenvectors_0 + \euler^{\masterequationeigenvalue_1 t}  c_1  \righteigenvectors_1  + \euler^{\masterequationeigenvalue_{1}^* t} c_{1}^* \Phi^{R^*}_{1}  ,
\label{eq:truncatedspectraldecomposition}
\end{align}
for $\dimension = 10^2,10^3$ at $\invtemperature = 4$.

\begin{figure}[h!]
\begin{center}
\includegraphics[width=0.47\textwidth]{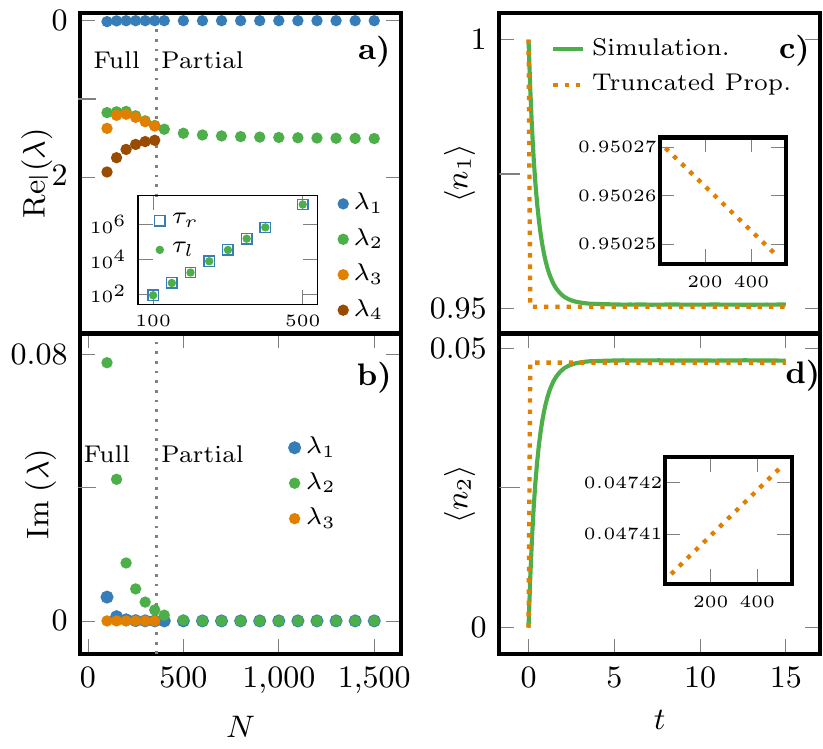}

\caption{Real a) and corresponding imaginary b) parts of the four most dominant eigenvalues with distinct and finite real part for different $\dimension$ and for $\invtemperature \!=\! 7$ as a representative of the AM phase. \comment{The inset in a) depicts the relaxation time scale $\tau_r$ and the lifetime of the metastable state $\tau_l$ as a function of $\dimension$.} Furthermore, the mean occupation density $ \langle \occupationdensity_i \rangle$ as a function of time for both the full (\ref{eq:spectraldecomposition}) and truncated (\ref{eq:truncatedspectraldecomposition}) propagation [$i$=1 in c) and $i$=2 in d)] for $\dimension=10^2$ is depicted. \label{fig:dominanteigenvaluesbetaseven} }
\end{center}
\end{figure}

To understand the metastability in the AM phase, Fig. \ref{fig:dominanteigenvaluesbetaseven} depicts the real and imaginary parts of the eigenvalues associated with the most dominant modes in panels a) and b), respectively, as a function of $\dimension$ for $\invtemperature =7$.
In contrast to Fig. \ref{fig:dominanteigenvalues}a), here, $\Real(\masterequationeigenvalue_2)$ clearly converges to a finite value with $\Real(\masterequationeigenvalue_1)$ quickly going to zero already for small $\dimension$. \comment{This is confirmed by the inset showing that $\tau_l$ and $\tau_r$ take very large values already for smaller systems implying that the metastability in the AM phase is much stronger than in the SM phase.}
As expected, in compliance with the nonoscillatory MF solution, the small magnitudes of the imaginary part vanish rapidly with growing system size as displayed in panel \ref{fig:dominanteigenvaluesbetaseven}b). Figs. \ref{fig:dominanteigenvaluesbetaseven}c) - d) reaffirm that the metastable state in the AM phase is reached at short time-scales and is quasistationary. Moreover, we note the large time-scales (cf. the scale of the axis of the insets) over which the metastable state can be observed in the dynamics in compliance with the observations made in panel \ref{fig:dominanteigenvaluesbetaseven}a).

Thus, we conclude from the observations made in this section, that for sufficiently large systems in the SM and AM phase at times $ \tau_m \! \ll \! t \! \ll \! \tau_r  $, the relaxation dynamics is determined by the metastable state associated with $\masterequationeigenvalue_{1,1^{*}}$ and the PFE. \comment{This time span is increasing with $\dimension$ [cf. Figs \ref{fig:dominanteigenvalues}a) and \ref{fig:dominanteigenvaluesbetaseven}a)] such that the metastable states can be observed over increasingly larger times. Owing to the PFT, any finite system will eventually leave these metastable states at times $t \gg \tau_r$ and relax into the unique stationary state at infinite time.} To sum up, we obtain the important result that the different phases and bifurcations of the MF dynamics are encoded in the spectrum of the Markov generator.

\section{Simulations}
\label{sec:simulations}

Solving the ME (\ref{eq:macromasterequation}) for systems on the order of {$\dimension \! \sim \! 10^3$} via full diagonalization of the propagator is computationally not feasible \footnote{We mention that the non-symmetric real matrix implies in general a complex eigensystem which shall be determined with float precision. The amount of random-access memory (RAM) required to diagonalize a matrix of dimension 80601 {$\times $} 80601 corresponding to a system consisting of 400 units is about 312 GB. We restrict the diagonalization to that size and employ different numerical methods for larger systems.}.
Hence for extremely large systems we resort to a stochastic simulation algorithm for computing the time evolution of the (Markov) jump processes. This dynamic Monte Carlo method, often referred to as Gillespie algorithm \cite{gillespie1976jcp,gillespie1977jcp}, generates trajectories of a stochastic process that are exact solutions to the stochastic process. By generating sufficiently many trajectories one can infer the statistics of the observables of the stochastic process, in particular the average values generically denoted by $\langle \cdot \rangle $.

\begin{figure}[h!]
\begin{center}

\includegraphics[width=0.48\textwidth]{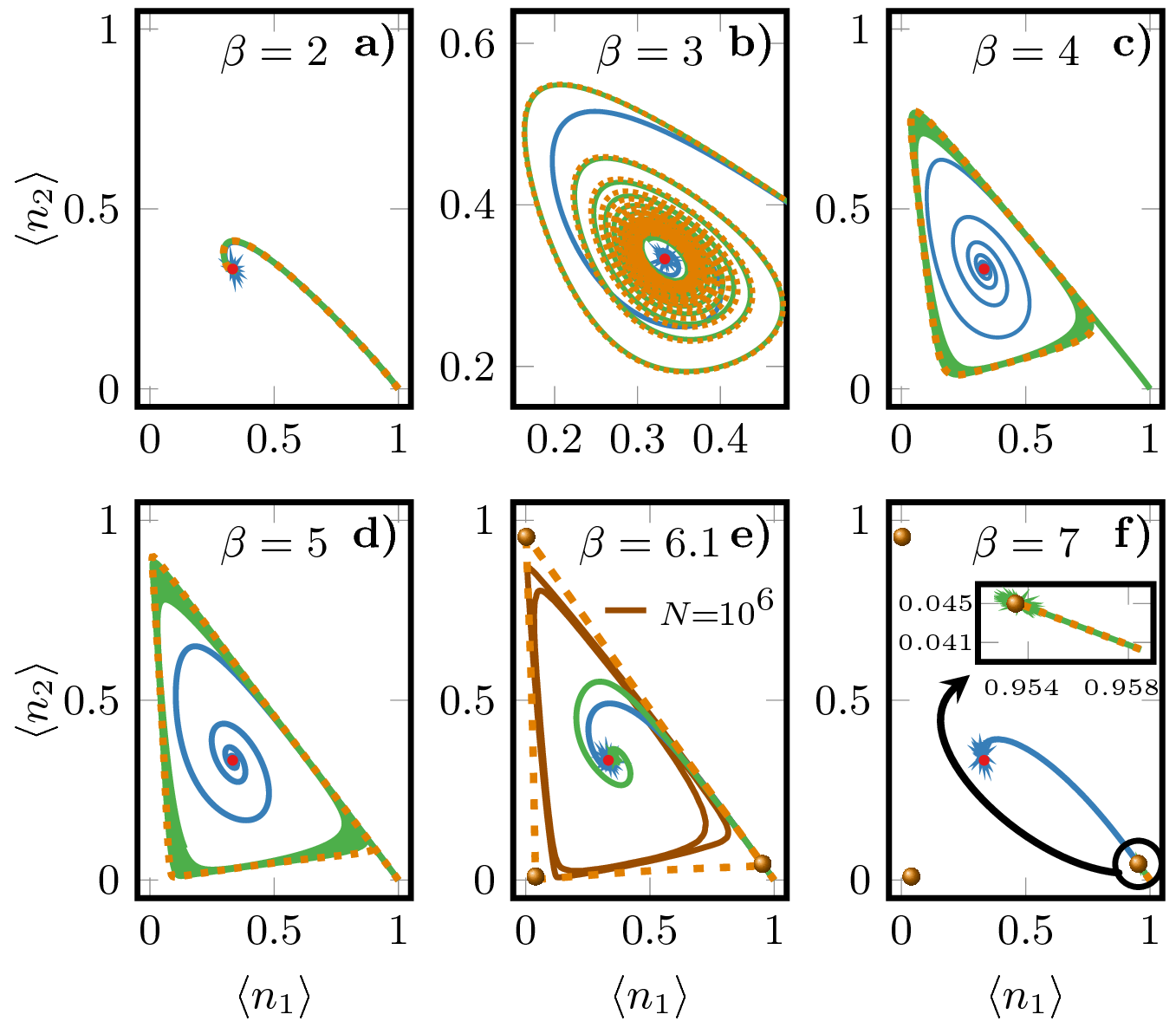}

\caption{Parametric plot of the mean occupation densities $\langle \occupationdensity_i \rangle$ for different finite system sizes $\dimension \!=\! 10^2$ (blue solid line), $\dimension\!=\!10^4$ (green solid line), and the MF limit $(\dimension \!=\! \infty$, orange dotted line) at distinct values of $\invtemperature$. In all panels we initialize the system in the ground state with $\occupationdensity_1=1$ and sample $10^6$ trajectories. \label{fig:stochasticdynamics} }
\end{center}
\end{figure}

Figure \ref{fig:stochasticdynamics} depicts the $\left\langle \occupationdensity_2 \right\rangle \! - \! \left\langle \occupationdensity_1 \right\rangle$ plots generated with the Gillespie algorithm sampling over $10^6$ trajectories for selected values of $\invtemperature$ and for different system sizes $\dimension = 10^2,10^4$. Except for $\invtemperature\!=\!6.1$ shown in e), the larger system, $\dimension\!=\!10^4$, agrees well with the MF limit at the displayed times. The smaller system, $\dimension\!=\!10^2$, significantly deviates in both the SM phase ($\invtemperature=4,5,6.1$) and AM phase ($\invtemperature=7$). In the A phase ($\invtemperature=2,3$), there are no visible differences between the different finite system sizes and the MF limit solution, as all are relaxing into the unique symmetric fixed point [red closed circle in panel a)]. 
Of particular interest is the dynamics for $\invtemperature = 7$. While the smaller system directly goes to the stationary state, the larger system quickly approaches and wiggles around the FP of the MF limit. This can be seen from the inset that displays a magnification around one of the MF FP [orange closed circle in f)]. Depending on the initial condition the metastable state will approach one of the three MF FPs.

This shows that the stochastic dynamics of sufficiently large systems indeed reproduces the MF dynamics at long times and thus confirms all predictions made above based on the spectral analysis.
As an exception, we observe in e) that close to the infinite-period bifurcation, $ \invtemperature \approx \invtemperature_{c_2} $, the large system does not exhibit the characteristics of the solution in the MF limit. However, an even larger system, $\dimension\!=\!10^6$, shows signatures of the LC albeit still deviating. These deviations are due to the strong fluctuations in the vicinity of the phase transition calling for larger $\dimension$ such that the finite system can accurately represent the deterministic limit.
We remark that this feature is also manifested in the increasing deviations between the LC frequency, $\limitcyclefrequency$, and the imaginary part of the crucial eigenvalue, $\masterequationeigenvalue_1$, as the second critical point, $\invtemperature_{c_2}\approx 6.1068$ is approached [cf. Fig. \ref{fig:eigenvaluesvsbeta}c)].

However, there is a set of initial conditions for which the stochastic dynamics will not go to one of these metastable states.
This set of initial conditions is readily constructed via all possible linear combinations of right eigenvectors of the mesoscopic generator from Eq. (\ref{eq:macromasterequation}), $ \macroprobabilityvector(0)  \!=\! \sum_{i\neq 1} a_i \, \righteigenvectors_i $, excluding the mode associated with the crucial eigenvalue pair $\masterequationeigenvalue_{1,1^*}$. It follows from the orthonormal dual-basis property of the eigensystem that the weights $c_{1,1^*}=0 $ in Eq. (\ref{eq:spectraldecomposition}). Hence the metastability would be removed from the dynamics.

\begin{figure*}
\begin{center}
\includegraphics[width=0.85\textwidth ,height=9cm]{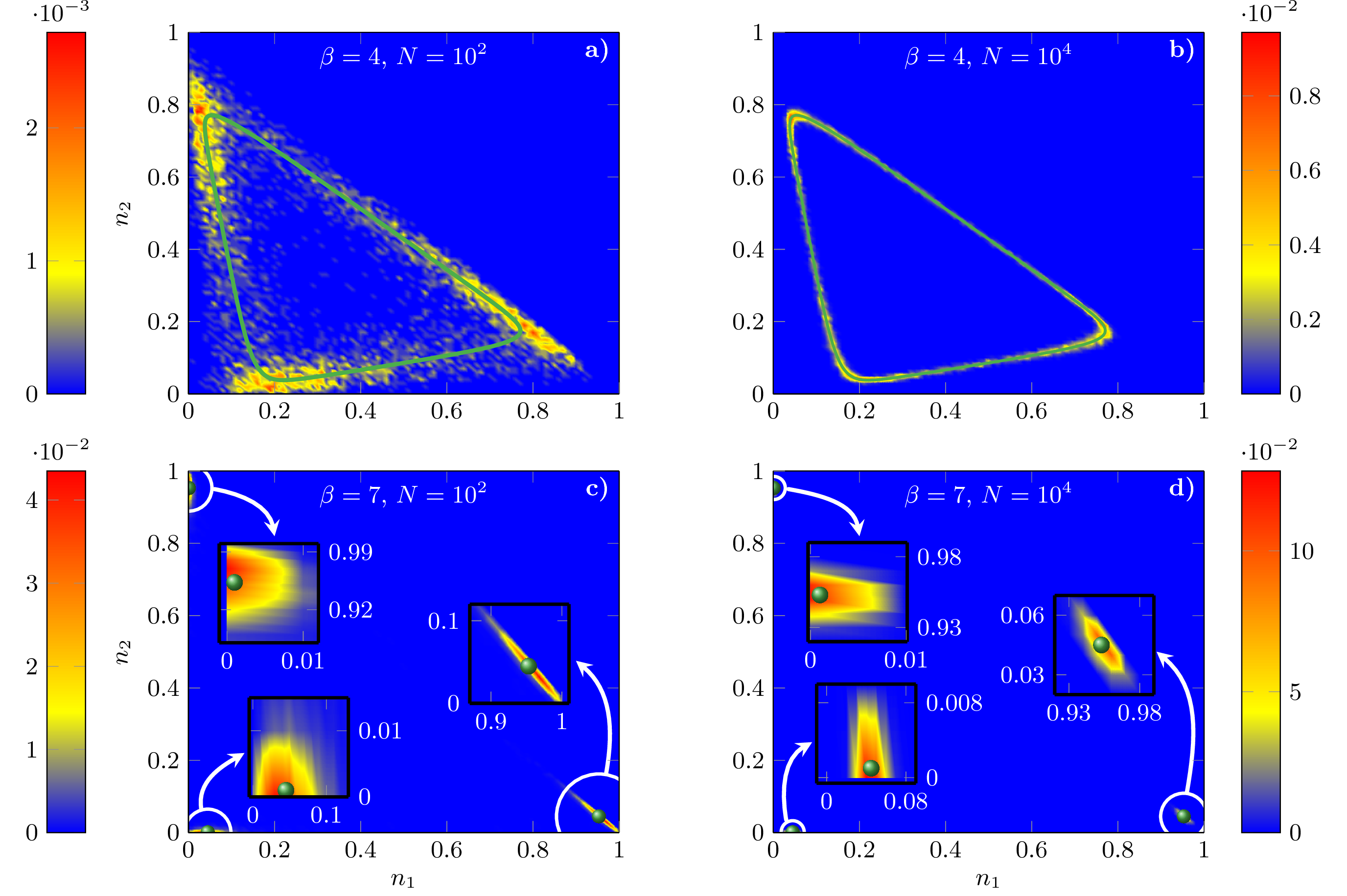}

\caption{Joint probability distribution $\jointprobability \left\lbrace \occupationdensity_1(t),\occupationdensity_2(t) \right\rbrace$ at $\invtemperature = 4$ in a), b) and at $\invtemperature = 7$ in c), d) for system sizes $\dimension = 10^2,10^3$ at time $t=~20$. The plots were created using a grid of dimension $101 \times 101$ that specifies the set of initial conditions. For comparison, in all plots the long-time MF solution (green solid line above, green closed circles below) is overlaid.
\label{fig:initialconditiondistribution}}
\end{center}
\end{figure*}

This prompts the question whether the metastability is a generic (up to a negligibly small set of special initial conditions) property of the stochastic process or just an artifact of choosing suitable initial conditions. This question is addressed in Fig. \ref{fig:initialconditiondistribution}, where the initial conditions are sampled and the joint probability distribution $\jointprobability\left\lbrace\occupationdensity_1(t=20),\occupationdensity_2(t=20) \right\rbrace$ for different system sizes $\dimension=10^2,10^4$ and $\invtemperature = 4,7$ is shown in a density plot.
In panel \ref{fig:initialconditiondistribution}a) the distribution exhibits its maxima indicated by the red spots close to the corners of the LC in the MF limit. Overall, the distribution clearly exhibits signatures of the LC but the probability mass is still dispersed around the LC contour. Moreover, over the entire state space there are regions with finite probability. If the system size is notably increased to $\dimension = 10^4$, as depicted in Fig. \ref{fig:initialconditiondistribution}b), the probability mass is sharply concentrated on the LC contour.

Turning to panels c) and d) corresponding to the AM regime with $\invtemperature=7$, we observe that the joint probability distribution for the smaller system already reproduces to a good approximation the three non-symmetric FPs in the MF limit. The distribution for the larger system further concentrates the probability mass on the three FPs as can be seen by comparing the insets on the left and on the right magnifying the vicinity of the FPs.
The convergence of the probability distribution at smaller $\dimension$ to the MF limit for larger $\invtemperature$ is consistent with the observations already made in the spectral analysis in Fig. \ref{fig:eigenvaluesvsbeta}.
We thus confirm, once again, that the metastability and therefore the convergence to the MF limit increases with $\dimension$ and $\invtemperature$.
Next, and more importantly, the emergence of the metastable state(s) is, up to a negligible set of special initial conditions, indeed a generic property of the stochastic process.
It is insightful to monitor the time evolution of $\jointprobability\left\lbrace\occupationdensity_1(t),\occupationdensity_2(t) \right\rbrace$ starting from a uniform grid at $t=0$ up to a time as the distribution becomes stationary or time-periodic.
To this end, the supplementaries \footnote{See supplementary media on \url{https://doi.org/10.6084/m9.figshare.5822097} } include movies displaying the dynamics of the distributions shown in Fig. \ref{fig:initialconditiondistribution}.

We have so far established a connection between linear stochastic dynamics and the deterministic nonlinear MF dynamics via the study of the spectrum of the Markov generator. Indeed, the different dynamical phases and bifurcations in the MF are encoded in the spectrum and appear as metastable states for long times in the stochastic dynamics. These predictions are confirmed by our simulations. We now proceed by analyzing the bifurcations as nonequilibrium phase transitions in the thermodynamic observables. In doing so, we link deterministic bifurcation theory to stochastic thermodynamics.

\section{Thermodynamic Laws}
\label{sec:thermodynamiclaws}

We first introduce the basic thermodynamic state functions in this model: the microscopic internal energy and the system entropy
\begin{subequations}
\begin{align} \label{eq:microinternalenergy}
\left\langle \microenergy \right \rangle &= \sum\limits_{\microstate} \macroenergy(\microstate) \, \microprobability_{\microstate} \\
\left\langle \microentropy \right \rangle &= - \sum\limits_{\microstate} \microprobability_{\microstate} \ln \microprobability_{\microstate} \, . \label{eq:microsystementropy}
\end{align}
\end{subequations}
For our setup with an autonomous driving, $\force$, these functions can only change due to the time-dependence of the probability distribution. The rate of change of internal energy
\begin{align}
\d_t \langle \microenergy \rangle = \sum\limits_{\microstate, \microstate'} \macroenergy(\microstate) \microrates_{\microstate \microstate'}  \, \microprobability_{\microstate'} 
= \langle \dot{\microheat} \rangle + \langle \dot{\microwork} \rangle , \label{eq:microfirstlaw}
\end{align}
naturally defines the microscopic first law of thermodynamics with the heat and work rate given by
\begin{align}
\langle \dot{\microheat} \rangle &= \sum\limits_{\microstate, \microstate'} \left[ \macroenergy(\microstate) - f \,\microsign  \right] \microrates_{\microstate \microstate'}  \, \microprobability_{\microstate'} \\
\langle \dot{\microwork} \rangle &= \sum\limits_{\microstate, \microstate'} f \; \microsign \; \microrates_{\microstate \microstate'}  \, \microprobability_{\microstate'} \, ,
\end{align}
where the sign function $\microsign$ is defined below Eq. (\ref{eq:fulltransitionrates}).
The microscopic local detailed balance relation (\ref{eq:microlocaldetailedbalance}) can be expressed in terms of the heat exchange with the bath along the forward transition
\begin{align} \label{eq:microlocaldetailedbalanceheat}
 \microheat(\microstate, \microstate')  =
- \frac{1}{\invtemperature} \ln \frac{\microrates_{\microstate \microstate'}}{\microrates_{\microstate' \microstate}} .
\end{align}
The system entropy change
\begin{align} \label{eq:microentropybalance}
\d_t \langle \microentropy \rangle = \langle \dot{\microentropy}_e \rangle + \langle \dot{\microep} \rangle
\end{align}
can be decomposed into the entropy flow from the bath to the system
\begin{align} \label{eq:microentropyflow}
\langle \dot{\microentropy}_e \rangle = -  \sum\limits_{\microstate, \microstate'} \microrates_{\microstate \microstate'} \, \microprobability_{\microstate'} \ln \frac{\microrates_{\microstate \microstate'}}{\microrates_{\microstate' \microstate}}
= \invtemperature \langle \dot{\microheat} \rangle ,
\end{align}
and the non-negative entropy production (EP) rate
\begin{align} \label{eq:microsecondlaw}
\langle \dot{\microep} \rangle =  \sum\limits_{\microstate, \microstate'} \microrates_{\microstate \microstate'} \, \microprobability_{\microstate'} \, \ln \frac{\microrates_{\microstate \microstate'} \microprobability_{\microstate'} }{\microrates_{\microstate' \microstate} \microprobability_{\microstate} } \geq 0 .
\end{align}
Equation (\ref{eq:microsecondlaw}) is the second law of thermodynamics and the inequality follows straightforwardly from $\ln x \leq x-1$.
The marginalization of the microscopic probability $ \microprobability_{\microstate} $ performed in Sec. \ref{sec:model}, yet being exact on the level of the dynamics, does not a priori guarantee that the thermodynamic observables defined above are invariant under this coarse-graining \cite{esposito2012pre}. Defining $\macroenergy_{\macrostate}$ to be the internal energy of the system in the macrostate $\macrostate$, and applying the coarse-graining from Eq. (\ref{eq:coarsegrainingidea}) on the expression for the internal energy in Eq. (\ref{eq:microinternalenergy}), we obtain
\begin{subequations}
\begin{align}
 \d_t \langle \microenergy \rangle
 &= \sum\limits_{\macrostate, \macrostate'} \, \sum\limits_{\microstate' \in \macrostate'} \, \sum\limits_{\microstate \in \macrostate} \macroenergy(\microstate) \, \microrates_{\microstate \microstate'} \, \microprobability_{\microstate'} \\
 &= \sum\limits_{\macrostate, \macrostate'} \marginalizedrates_{\macrostate \macrostate'} \, \macroenergy(\macrostate) \! \sum\limits_{\microstate' \in \macrostate'}  \microprobability_{\microstate'}  \sum\limits_{\microstate \in \macrostate}  1 \\
&= \sum\limits_{\macrostate, \macrostate'} \macroenergy(\macrostate) \, \macrorates_{\macrostate \macrostate'} \, \macroprobability_{\macrostate'} \equiv  \d_t \langle \macroenergy \rangle .
\end{align}
\end{subequations}
Thus the coarse-graining admits a representation in the mesospace while it keeps the internal energy invariant. The heat and work fluxes can also be exactly coarse-grained as
\begin{subequations}
\begin{align} \label{eq:macroheat}
\langle\dot{\macroheat}\rangle &=  \sum\limits_{\macrostate, \macrostate'}  \underbrace{ \big( \macroenergy(\macrostate) - f \, \macrosign  \big) \macrorates_{\macrostate \macrostate'}  \macroprobability_{\macrostate'}  }_{ \macroheat(\macrostate, \macrostate') } \\
\langle\dot{\macrowork}\rangle &= \sum\limits_{\macrostate, \macrostate'} f \, \macrosign  \macrorates_{\macrostate \macrostate'} \, \macroprobability_{\macrostate'}  . \label{eq:macrowork}
\end{align}
\end{subequations}
Consequently, the first law of thermodynamics has a closed mesoscopic representation which is identical to the one from Eq. \eqref{eq:microfirstlaw}.
%\begin{align}
%\d_t \left\langle \macroenergy \right\rangle = \langle \dot{\macroheat} \rangle + \langle \dot{\macrowork} \rangle . \label{eq:macrofirstlaw}
%\end{align}
We note that after the coarse-graining the heat increment
\begin{subequations}
\begin{align} \label{eq:macrolocaldetailedbalanceheat}
\!\!\!  \macroheat(\macrostate, \macrostate') 
&\!=\! - \frac{1}{\invtemperature} \ln \frac{\macrorates_{\macrostate \macrostate'}}{\macrorates_{\macrostate' \macrostate}} \!+\! \frac{1}{\invtemperature} \, \change \internalentropy(\macrostate,\macrostate') \\
&= - \frac{1}{\invtemperature} \ln \frac{\tilde{\macrorates}_{\macrostate \macrostate'}}{\tilde{\macrorates}_{\macrostate' \macrostate}} ,
\end{align}
\end{subequations}
is no longer directly given by the local detailed balance relation like in the microspace, cf. Eq. (\ref{eq:microlocaldetailedbalanceheat}), but also contains the internal entropy from Eq. (\ref{eq:internalentropy}) \cite{esposito2012pre}. 
We define the system entropy in the mesospace
\begin{align}  \label{eq:macrosystementropy}
\left\langle \macroentropy \right \rangle &=  \sum\limits_{\macrostate} \macroprobability_{\macrostate} \left( \multiplicity(\macrostate) - \ln \macroprobability_{\macrostate} \right) \, ,
\end{align}
consisting of the non-equilibrium entropy defined by Eq. (\ref{eq:microsystementropy}) and the internal entropy accounting for the multiplicity of distinct microscopic configurations for a given macrostate. Analogously to Eq. \eqref{eq:microentropybalance}, we decompose the time-derivative of the entropy into the entropy flow 
\begin{align} \label{eq:macroentropyflow}
\langle \dot{\macroentropy}_e \rangle = -  \sum\limits_{\macrostate, \macrostate'} \macrorates_{\macrostate \macrostate'} \, \macroprobability_{\macrostate'} \ln \frac{ \tilde{\macrorates}_{\macrostate \macrostate'}}{\tilde{\macrorates}_{\macrostate' \macrostate}}
= \invtemperature \langle \dot{\macroheat} \rangle ,
\end{align}
and the EP rate
\begin{align} \label{eq:macrosecondlaw}
\langle \dot{\macroep} \rangle =  \sum\limits_{\macrostate, \macrostate'} \macrorates_{\macrostate \macrostate'} \, \macroprobability_{\macrostate'} \, \ln \frac{\macrorates_{\macrostate \macrostate'} \macroprobability_{\macrostate'} }{\macrorates_{\macrostate' \macrostate} \macroprobability_{\macrostate} } \geq 0 .
\end{align}
The definitions in Eqs. \eqref{eq:macrosystementropy},\eqref{eq:macrosecondlaw} are in general not coinciding with those made at the microscopic level, \idest $\langle \macroentropy \rangle \neq \langle \microentropy \rangle, \langle \macroep \rangle \neq \langle \microep \rangle $. The nonlinearity of the system entropy and the EP [Eqs. (\ref{eq:microsystementropy}), \eqref{eq:microsecondlaw}] in the microstate probability $\microprobability_{\microstate}$ is incompatible with the coarse-graining. Instead, an application of Eq. (\ref{eq:microsecondlaw}) gives rise to additional entropic contributions which are dependent on microscopic information, hence the coarse-grained equation can not be closed \cite{esposito2012pre}. For the special case of a stationary probability distribution, $\macrosteadyprobability_{\macrostate}$, one can show (cf. appendix \ref{sec:spanningtreeproof}) via the spanning tree formula \cite{schnakenberg1976rmp} that the microstates belonging to the same macrostate are equally probable, $ \microprobability_{\microstate} =  \macroprobability_{\macrostate} / \multiplicity(\macrostate) $. In the stationary limit, the entropies in mesoscopic representation are therefore identical to those in microscopic representation, \idest $\langle \macroentropy^{s} \rangle = \langle \microentropy^{s} \rangle, \langle \macroep^{s} \rangle = \langle \microep^{s} \rangle $. For this particular case, the second law
\begin{align} \label{eq:macrosecondlawstationary}
\!\!\!\! 
\langle \dot{ \macroep^s } \rangle  \! = \! \sum\limits_{\macrostate, \macrostate'} \!\! \macrorates_{\macrostate \macrostate'} \, \macrosteadyprobability_{\macrostate'} \, \ln \frac{ \tilde{\macrorates}_{\macrostate \macrostate'}}{\tilde{\macrorates}_{\macrostate' \macrostate}} 
= - \langle \dot{\macroentropy}_e^s  \rangle \geq 0  ,
\end{align}
boils down to the steady entropy flow $\langle \dot{ \macroentropy }_e^s \rangle$ being equal to the magnitude of the steady EP rate $\langle \dot{ \Sigma }^s \rangle$. Using the non-positivity of the average stationary heat, $\invtemperature \langle \dot{ \macroheat }^s \rangle \leq 0 $, we easily verify that $\langle \dot{ \macroep }^s \rangle \geq 0$.

We now turn to the MF case and consistently define the first law in this limit
\begin{align}
\d_t \, \meanfieldenergy  = \sum\limits_{i,j} \macroenergy_{i} \, \meanfieldrates_{i j} \, \meanfieldprobability_{j}   
=  \dot{\mathcal{\macroheat}} +  \dot{\mathcal{\macrowork}}  ,
\end{align}
with the heat and work flux
\begin{subequations}
\begin{align}
\dot{\mathcal{\macroheat}} &= \sum\limits_{i,j} \big( \macroenergy_{i} - f \, \meanfieldsign \big) \, \meanfieldrates_{ij} \, \meanfieldprobability_{j}  \\
\dot{\mathcal{\macrowork}} &= \sum\limits_{i,j} f \, \meanfieldsign \, \meanfieldrates_{ij} \, \meanfieldprobability_{j} \, , \label{eq:meanfieldwork}
\end{align}
\end{subequations}
where $i,j = 1,2,3$ specifies the state of the single MF unit. In analogy to Eq. \eqref{eq:microsystementropy}, we write the system entropy in the MF limit as
\begin{align} \label{eq:meanfieldentropy}
 \meanfieldentropy &= - \sum\limits_{i} \meanfieldprobability_{i} \ln \meanfieldprobability_{i} ,
\end{align}
which we split into the MF entropy flow
\begin{align} \label{eq:meanfieldentropyflow}
\dot{\meanfieldentropy}_e = -  \sum\limits_{i, j} \meanfieldrates_{i j} \, \meanfieldprobability_{j} \ln \frac{\meanfieldrates_{i j}}{\meanfieldrates_{j i}}
= \invtemperature  \dot{\meanfieldheat}  ,
\end{align}
and the non-negative MF EP rate
\begin{align} \label{eq:meanfieldep}
\dot{\meanfieldep} =  \sum\limits_{i, j} \meanfieldrates_{i j} \, \meanfieldprobability_{j} \, \ln \frac{\meanfieldrates_{i j} \meanfieldprobability_{j} }{\meanfieldrates_{j i} \meanfieldprobability_{i} } \geq 0 .
\end{align}
As the MF represents the asymptotic limit of the mesospace, it holds that all the mesoscopic averages of the intensive observables $\langle X \rangle / \dimension $ that are consistent with the coarse-graining in Eq. (\ref{eq:coarsegrainingidea}) converge to the corresponding observables $\mathcal{X}$ in the MF limit, $\lim\limits_{\dimension \to \infty} \tfrac{\langle\dot{X}\rangle}{\dimension} = \dot{\mathcal{X}},$ with $X \!\! ~=~ \!\! \macroenergy,\macroheat,\macrowork,\macroentropyflow,\occupation_i $. Consequently, for the MF definitions in Eqs. \eqref{eq:meanfieldentropy} and \eqref{eq:meanfieldep} to represent the physical entropies, we have to restrict to the stationary case, $ \meanfieldsteadyprobability $, which yields for the second law in the MF limit
\begin{align} \label{eq:meanfieldsecondlaw}
\dot{ \meanfieldentropy }_i^s = \sum\limits_{i,j} \meanfieldrates_{ij} \, \meanfieldsteadyprobability_{j} \ln \frac{\meanfieldrates_{ij}}{\meanfieldrates_{ji}} = - \dot{ \meanfieldentropy }_e^s \geq 0  
\end{align}
The non-negativity of the MF EP follows from the non-positivity of the MF heat in this model.
We have thus developed three different levels (microspace, mesospace and MF) to consistently characterize the energetics of our model. For the first law, the lower levels of description are equivalent, while for the second law they only coincide in the stationary limit. The same applies asymptotically in the macroscopic limit to the thermodynamic observables defined at the MF level.

\section{Dissipated Work}
\label{sec:dissipatedwork}

With the thermodynamic framework developed in the preceding section at hand, we can now proceed by addressing one of the crucial research questions of this work, that is the thermodynamics of non-equilibrium phase transitions. We are naturally interested in the (metastable) synchronization regime bounded by the two phase transitions. Since the nonstationary EP represented in the microscopace is not identical to the one in the mesospace [Eqs. \eqref{eq:microsecondlaw} and \eqref{eq:macrosecondlaw}], we characterize the nonequilibrium phase transitions via the dissipated work given by Eqs. (\ref{eq:macrowork}) and \eqref{eq:meanfieldwork}. At metastable or infinite time, the work is observed to be always dissipative on average, that is the system takes up the energy from the nonconservative force, $\langle \macrowork \rangle > 0 $, and dissipates it into the bath in the form of heat, $\langle \macroheat \rangle < 0 $, for all temperatures and system sizes.

\begin{figure}[h!] 
\begin{center}

\includegraphics[width=0.48\textwidth]{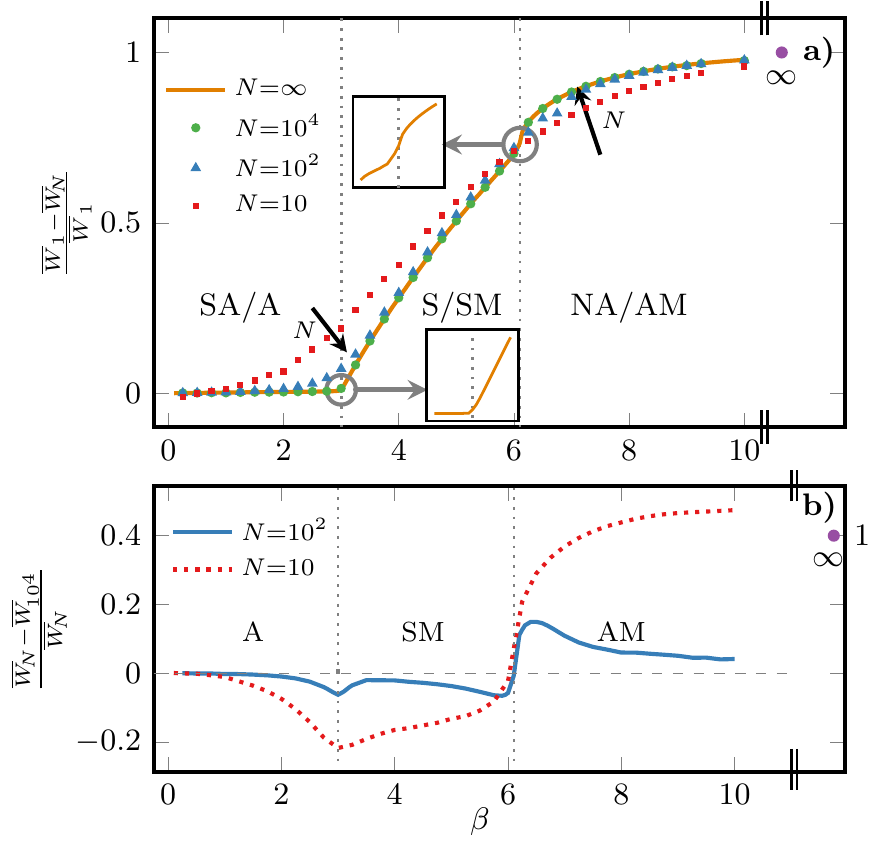}
\includegraphics[width=0.48\textwidth]{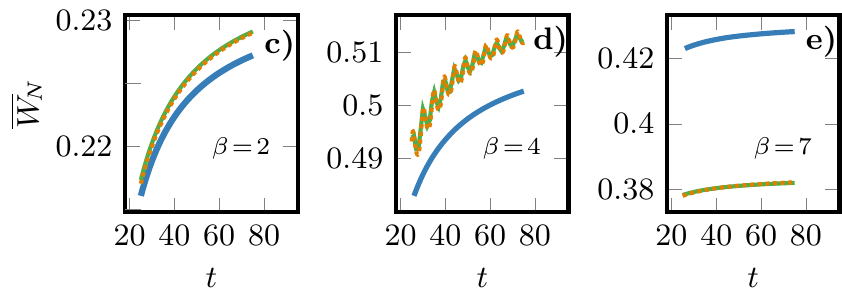}

\caption{Panel a): Difference of the dissipated work for a single-unit, $ \singleworkrate $, and for a unit in a network of size $\dimension$, $ \overline{\macrowork}_{\dimension} $, for inverse temperatures $\invtemperature = 0 \ldots 10$. The time $t=500$ is chosen to ensure that $ \overline{\macrowork}_{\dimension} $ has converged to its (metastable) asymptotic value.
Panel b): Difference of the dissipated work per unit for networks of different size with $\dimension < 10^4$, for $\invtemperature$ ranging from 0 to 10 and thus covering all three phases: 
Symmetric asynchronous phase (SA), synchronous phase (S), non-symmetric asynchronous phase (NA) in the MF and asynchronous phase (A), synchronous metastable phase (SM), and the asynchronous metastable phase (AM) for finite metastable systems. As in panel a), the time is $t=500$. The purple closed circles in panels a) and b) represent the analytic expression given by Eqs. (\ref{eq:finitesinglework_largebeta}) and (\ref{eq:finiteworkasymptotic_largebeta}), respectively. Panels c)--e): Plot of $\overline{\macrowork}_{\dimension}$ for selected values at $\invtemperature = 2,4,7$ and system sizes $\dimension \!=\! 10^2$ (blue solid line) and $\dimension \!=\! 10^4$ (green solid line). This is the same data as the one underlying the blue solid curve in plot b) but, for better visualization, the time $t$ is restricted from 20 to 80. For comparison, the MF limit (orange dashed lines) is overlaid in c)--e). In each plot all finite systems were simulated sampling $10^6$ trajectories. \label{fig:dissipatedwork}  }
\end{center}
\end{figure}

Figure \ref{fig:dissipatedwork}a) depicts the difference between the stationary work current of a single unit, $ \singleworkrate = 2 \, \arrheniusprefactor  \force \sinh \! \left( \force \invtemperature /2 \right) $, and the asymptotic work current per unit in a network of size $\dimension$, $ \overline{\macrowork}_{\dimension} \! \equiv \! \frac{\langle \macrowork \rangle}{Nt}$ as a function of $\invtemperature$ for different $\dimension$. The derivation of the single-unit stationary work current, $\singleworkrate$, is deferred to appendix \ref{sec:singleunitproof} and given by Eq. \eqref{eq:singleunitworkcurrent}. The asymptotic work current, $ \overline{\macrowork}_{\dimension}$, is numerically determined by solving Eqs. (\ref{eq:macromasterequation}) and (\ref{eq:meanfieldmasterequation}) for a finite and a MF system, respectively. As seen in Fig. \ref{fig:dissipatedwork}a), the large ($\dimension \! = \! 10^4$) system agrees excellently with the MF limit for all temperatures, while the smaller systems, albeit showing a qualitatively similar behavior, unlike the dynamics, deviate significantly.

Since the single unit work current is governed by a smooth and convex function, we observe that the dissipated MF work exhibits striking changes at the critical points $\invtemperature_{c_{1,2}}$. The vicinities of these critical points are magnified in the two insets. The phase transitions in the dissipated MF work at $\invtemperature_{c_1}$ and $\invtemperature_{c_2}$ exhibit a kink and a saddle, respectively, and are therefore reminiscent of a first- and second-order equilibrium phase transition. Remarkably, owing to the metastability in the stochastic dynamics, sufficiently large systems also exhibit finite-time signatures of these nonequilibrium phase transitions at the bifurcation points which blur out with decreasing system size.

In the high-temperature limit, $\invtemperature \to 0$, the difference $ \singleworkdifference ~ \equiv ~ \singleworkrate ~-~ \overline{\macrowork}_{\dimension}$ between the dissipated work of a single unit and an interacting system per unit is always zero since the interaction energy gets canceled ($\occupation_i = \occupation_j$ in Eq. (\ref{eq:changeinternalenergy}) and $\potential/\dimension \to 0$ as $\dimension \to \infty$.). While for the MF this holds true in the entire SA phase, for finite systems the range of $\invtemperature$ values in the A phase for which the interaction energy is negligible decreases with $\dimension$. We find that interactions reduce the costs to maintain the system in its nonequilibrium state, $\singleworkdifference \!>\! 0$. This work dissipation gap, $\singleworkdifference$, is a monotonically increasing function of $\invtemperature$ and becomes infinitely large in the low-temperature limit, since $\singleworkdifference / \singleworkrate \to 1$ as $\invtemperature \to \infty$.
This asymptotic limit can be seen as follows. In Sec. \ref{sec:meanfielddynamics} we observed that in the low-temperature limit, one can make use of the equilibrium picture where the system tends to occupy its energy ground states. In this limit, we have for the dissipated work of a finite network per unit 
\begin{align}
\lim_{\invtemperature \to \infty} \overline{\macrowork}_{\dimension} =
\lim_{\invtemperature \to \infty} \arrheniusprefactor  \force \left(\euler^{\invtemperature  \force}-1 \right) \euler^{-\frac{\invtemperature  (\force \dimension - \dimension \potential + \potential)}{2 \dimension} } ,
\end{align}
which is subdominant to $\singleworkrate$ [cf. Eq. \eqref{eq:singleunitworkcurrent}]
\begin{align} \label{eq:finitesinglework_largebeta}
\lim\limits_{\invtemperature \to \infty} \frac{ \Delta \overline{\macrowork} }{\singleworkrate} = 	\lim\limits_{\invtemperature \to \infty}  1- \euler^{ \frac{\invtemperature \potential (\dimension -1)}{2 \dimension} } = 1 .
\end{align}

Hence we have shown that at low and intermediate temperatures an interacting network of any size is energetically favorable with respect to a noninteracting one. Interestingly, in the the two phases of higher temperature, the operational costs per unit can be further decreased by employing \emph{smaller} networks. As one approaches the second critical point the different curves intersect and in the NA/AM phase the operation of \emph{larger} networks gives rise to less work dissipation per unit.

This is also illustrated in Fig. \ref{fig:dissipatedwork}b) that depicts the difference in the dissipated work between a system of size $\dimension = 10^4$ exhibiting metastability and a smaller system which does not display metastable states. In agreement with panel a), the smaller system requires less input per unit to be maintained in the two higher temperature phases, since the difference $ \finiteworkdifference \equiv \overline{\macrowork}_{\dimension} - \overline{\macrowork}_{10^4} < 0 $ while the opposite holds true in the AM phase, where $ \finiteworkdifference > 0 $.

Again, we observe at the critical points significant changes in $ \finiteworkdifference $: At the first critical point $\finiteworkdifference$ takes a local minimum and at the second critical point it changes sharply around an inflection point. It is plausible that these changes are more pronounced for decreasing $\dimension$ as the reference system ($\dimension'=10^4$) exhibits metastability, such that for increasing differences in the network size compared the distance to metastable behavior implying signatures of phase transitions in the dissipated work becomes larger.

For the same reasons as stated in the context of plot \ref{fig:dissipatedwork}a), $ \finiteworkdifference $ goes to zero in the high-temperature regime, while in the low-temperature limit one obtains
\begin{align} \label{eq:finiteworkasymptotic_largebeta}
\lim\limits_{\invtemperature \to \infty} \frac{ \finiteworkdifference }{\overline{\macrowork}_{\!\! \dimension}} = \lim\limits_{\invtemperature \to \infty}  1- \euler^{-\frac{\invtemperature \potential}{2} \left( \frac{1}{10^4} - \frac{1}{\dimension} \right)} = 1,
\end{align}
if $ \dimension < 10^4 $. This limit is illustrated by the purple closed circle in the plot. For the larger system the work difference is decreasing in the range of available data. Generating data for larger $\invtemperature$ to monitor the convergence to the low-temperature limit is not possible since the simulation becomes numerically unstable owing to the large values the exponentials take in the transition rates.

To illustrate the data underlying the plots in Fig. \ref{fig:dissipatedwork}b), we show in panel c) to e) the time-scaled work asymptotics per unit for different system sizes as chosen for the blue curve in panel b) as well as the MF limit for selected values of $\invtemperature=2,4,7$. We note the excellent agreement between the MF limit and the large system in compliance with the observations made in panel \ref{fig:dissipatedwork}a). On the other hand, the small system clearly deviates from the large systems in all three different regimes, even though we observed that in the SA/A phase the dynamics of large and small systems can hardly be distinguished. Due to the approximate time-periodicity in the S/SM phase, the dissipated work is also oscillating.

\begin{figure}[h!] 
\begin{center}

\includegraphics[scale=1]{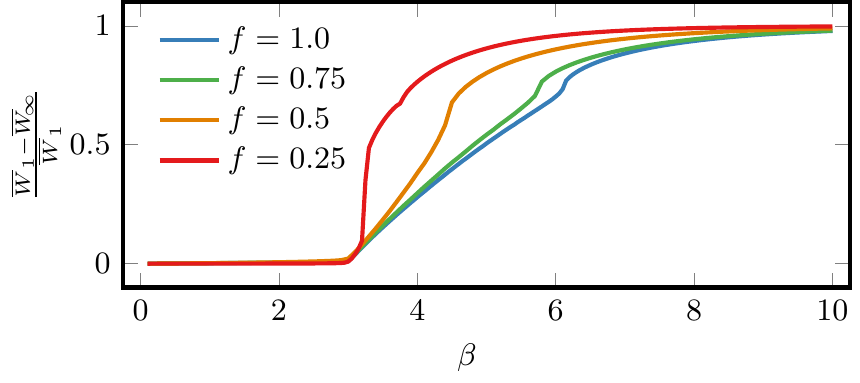}

\caption{\comment{Comparison between stationary single-unit work current $\singleworkrate$ and asymptotic MF work current $\overline{\macrowork}_{\!\! \infty}$ as a function of $\invtemperature$ for $\force=0.25,\,0.5,\,0.75,\,1.0$. The time $t=500$ is chosen such that the time-averaged MF work has converged to its asymptotic value.} \label{fig:meanfieldwork}}
\end{center}
\end{figure}

\comment{Finally, Fig. \ref{fig:meanfieldwork} depicts the difference between the stationary single-unit and the asymptotic MF unit work current, $\Delta \overline{\macrowork}_{1 \infty}$, as a function of $\invtemperature$ for different $\force$. Again, $ \Delta \overline{\macrowork}_{1 \infty} =0 $ in the A phase since the single and the MF unit are indistinguishable in the high-temperature regime as shown above in the context of Fig. \ref{fig:dissipatedwork}a). For $\beta \geq \beta_{c_1}$ the second critical point is gradually shifting to smaller $\beta$ [cf. Fig. \ref{fig:meanfieldparameterphasespace}] while the difference $ \Delta \overline{\macrowork}_{1 \infty} $ is monotonically increasing with decreasing $f$. Therefore, if compared to the MF, the additional costs to maintain the nonequilibrium stationary state of the noninteracting system at a given temperature are the smaller the further it is driven out-of-equilibrium. This implies in particular that the dissipation of the synchronized system at fixed temperature is approaching the one of the non-synchronized system as they are further driven out-of-equilibrium.
}

To summarize, we have obtained two major results in this section. 
First, though the nonequilibrium phase transitions are naturally only present in the MF-limit where the nonlinear dynamics exhibits the supercritical Hopf and the infinite-period bifurcation, we find that the metastability observed in the finite-system dynamics translates into signatures of genuine nonequilibrium phase transition. This consistently connects linear stochastic dynamics, nonlinear deterministic dynamics, and thermodynamics and furthermore demonstrates that thermodynamics of nonequilibrium phase transitions and bifurcation theory are closely related. Secondly, any finite and attractive interaction in a network reduces the dissipated work per unit. Interestingly, if operating in the synchronous phase, it is even more economic to employ interacting but \emph{smaller} networks. What is still open to investigate is how the nonequilibrium phase transitions affect the power-efficiency trade-off, if the system operates as an energy-converting machine.

\section{Efficiency at maximum power}
\label{sec:efficiencyatmaxpower}

In order to construct such an energy converter with our system both a positive force $\inforce > 0$ and a negative force $\outforce < 0$ are applied on the same unit. Examples for this type of work-to-work conversion are could be double quantum dot channel capacitively coupled to a quantum point contact \cite{bulnes2015njp} or the biological motors kinesin and myosin. In the latter case, the motor is driven forward with $\inforce$ by extracting energy via ATP hydrolysis while the load carried by the motor is modeled as $\outforce$ \cite{vale2000science,imparato2015njp}. In general, these two forces obey two different distributions accounting for the crucial fluctuations these motors exhibit.
Since the following discussion is restricted to the MF limit, we consider the homogenous case where the same positive and negative force are applied on all units.
%We remark that homogenous motors modeled as diffusing particles on a lattice subjected to an exclusion rule were studied in \cite{imparato2012prl} while a noisy Kuramoto model resembling molecular motors with nontrivial force distributions was investigated in \cite{imparato2015njp}.

We thus decompose the net force $\force = \inforce + \outforce$ into the driving force $\inforce >0$ and the load force $\outforce < 0$. Their respective steady-state work contributions are denoted by $ \meanfieldwork_1^{s} $ and $ \meanfieldwork_2^{s} $. Substituting Eq. (\ref{eq:meanfieldlocaldetailedbalance}) into Eq. (\ref{eq:meanfieldsecondlaw}), yields the following decomposition of the stationary EP in the MF limit
\begin{align} \label{eq:entropyproductiondecomposition}
\meanfieldep^s = \meanfieldep^{\meanfieldwork_1^s} + \meanfieldep^{\meanfieldwork_2^s} \, ,
\end{align} 
where $\meanfieldep^{\meanfieldwork_k^s} \!=\! \invtemperature \, \meanfieldwork_k^s, \, k \!=\! 1,2$.
Based on Eq. (\ref{eq:entropyproductiondecomposition}), we use as an unambigious definition for the efficiency of this work-to-work conversion (cf. Refs. \cite{espositovandenbroeck2012pre,verley2014nc})
\begin{align}
\efficiency = - \frac{ \meanfieldep^{\meanfieldwork_2^s} }{ \meanfieldep^{\meanfieldwork_1^s} } 
% = - \frac{\outforce}{\inforce} 
= 1 - \frac{\force}{\inforce}  . \label{eq:efficiency}
\end{align}
At equilibrium ($\force \!=\! 0$), the reversible limit, $ \efficiency_{c}\!=\!1 $ is attained while out of equilibrium ($\force \neq 0$) the efficiency is bounded, $0 \!<\! \efficiency \!<\! 1$.
Of particular interest is the efficiency at maximum power (EMP) \cite{curzon75ajp}, which results from the optimization of the stationary output power $\meanfieldpower \equiv \D \meanfieldwork_2^s/ \D t$ with respect to the output force
\begin{align}
\maxefficiency = 1 - \left. \frac{\maxforce}{\inforce} \right|_{\maxforce =\! \inforce \!-\! \maxoutforce} , \label{eq:maxefficiency}
\end{align}
The maximization parameter $\force^*$ is determined by the condition $ \D \meanfieldpower / \D \outforce \!=\! 0$, while fixing $\inforce \!=\! 1$ and thus varying the total dissipation.

\begin{figure}[h!] 
\begin{center}

\includegraphics[width=0.5\textwidth]{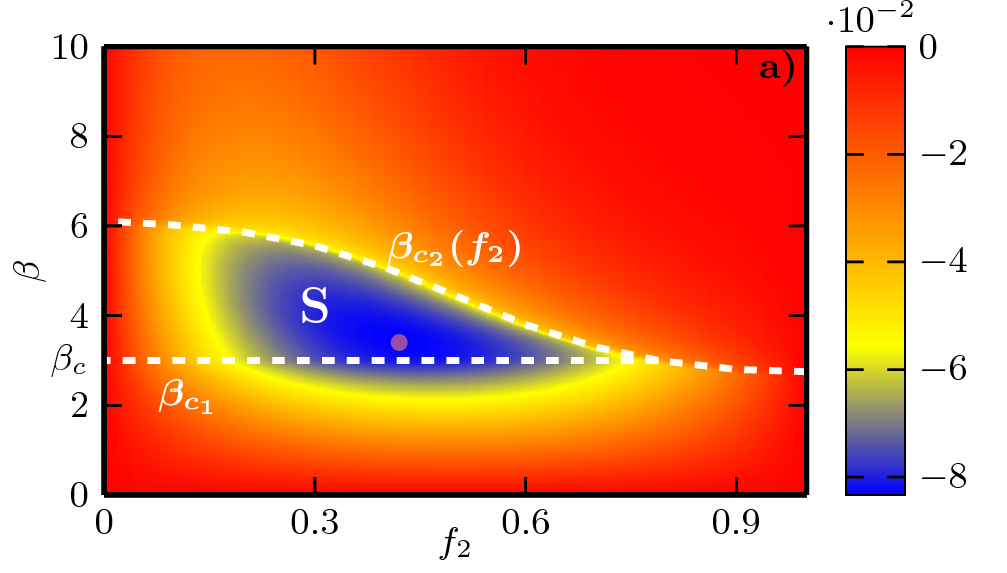}  \smallskip

\includegraphics[width=0.48\textwidth]{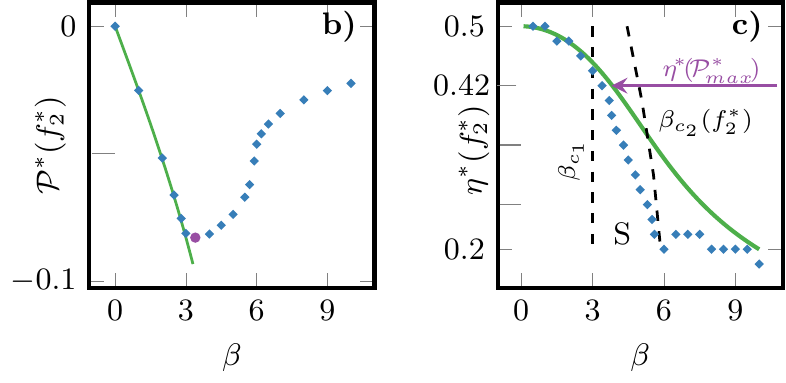}

\caption{Depiction of the output power a) as a function of the output force $\outforce$ and the inverse temperature $\invtemperature$. The white dashed lines correspond to the numerically determined critical points as a function of the output force. Hence the enclosed area defines the synchronization phase S. The global maximum of the output power is indicated by the purple closed circle.
In panel b) the maximum output power $\maxmeanfieldpower$ is optimized with respect to $\outforce$ and in panel c) the associated EMP $\maxefficiency(\maxoutforce)$ is displayed. In panel c) the dashed lines specify the critical points and the synchronization phase S. The efficiency at the global maximum power is indicated by the purple arrow.
The (semi-)analytic solution for $\invtemperature < \invtemperature_c$ [green lines] is overlaid with the numerical data in the lower panels.
 \label{fig:emp} }
\end{center}
\end{figure}

In the SA phase, $\invtemperature < \invtemperature_c$, the stationary power putput coincides with the average work current of a single unit given by Eq. \eqref{eq:singleunitworkcurrent}. For the other two phases (S and NA), we have to resort to simulations to obtain the power output. Moreover, owing to the time-periodic state in the S phase, the power is periodically changing in time. Hence we consider the time-average of the power over one LC period. Figure \ref{fig:emp}a) shows the numerically determined output power $\meanfieldpower$ as a function of $\invtemperature$ and $\outforce$ in a density plot.

The white dashed lines indicate the critical points $\invtemperature_{c_{1,2}}$ as a function of the output force. Thus the area enclosed by those lines corresponds to the S phase. Remarkably, we find that the maximum output power is generated in this phase. In particular, the global maximum of the output power indicated by the purple closed circle lies inside the S phase. At large $\invtemperature$ that represents the NA phase, the generated power rapidly drops. In panel b) the output power maximized with respect to the output force for different values of the inverse temperature is depicted. The numerical data from panel a) is overlaid with the (semi-)analytic results in the SA phase (green solid line) and the low-temperature limit and shows an excellent agreement. These limiting cases can be obtained as follows. In the SA phase, the condition for maximization of the power
\begin{align} \label{eq:lowbetapowerforce}
\frac{\D \meanfieldpower}{\D \outforce}  &= \left[ 2 + \invtemperature \outforce + \euler^{\invtemperature(1 - \outforce)} ( \invtemperature \outforce - 2) \right] = 0 ,
\end{align} 
results in a transcendental equation that must be treated numerically. In the low-temperature limit, the extremum condition
\begin{align}  \label{eq:largebetapowerforce}
\frac{\D \meanfieldpower}{\D \outforce}  &= \! 
\euler^{\frac{\invtemperature}{2} ( \outforce - 1)} \! \left[ \euler^{\invtemperature(1 - \outforce)} ( \invtemperature \outforce \!- \! 2) \!+\! ( \invtemperature \outforce \!+ \! 2) \right] \! = 0 ,
\end{align}
can not be satisfied for any $\outforce$ compatible with the constraint $\invtemperature = \infty$.

The efficiencies associated with the processes corresponding to the data points in panel \ref{fig:emp}b) are depicted in panel c). Again, the semianalytic solution for the temperatures corresponding to the SA phase (green solid line) is compared with the numerical results and shows an excellent agreement at these temperatures. As $\invtemperature$ approaches zero, the EMP takes the universal linear-response value for tightly-coupled (only one net-current) systems, $ \maxefficiency = 0.5 \, \efficiency_{c} $ \cite{vandenbroeck2005prl,esposito2009prl}. This can be seen by expanding the expression for the stationary work current in the SA phase given by Eq. (\ref{eq:singleunitworkcurrent}) up to first order in $\invtemperature$ which yields the linear-response relation $J^s \! \approx \! L \, \force $ with the Onsager coefficient $L = \arrheniusprefactor \, \invtemperature $. Therefore, small products $\invtemperature \force$ correspond to linear response in our model and lead to EMP values very close to 1/2. With increasing $\invtemperature$, the system starts to respond nonlinearly and the efficiency decreases monotonically and nonlinearly.

It is worth emphasizing that the efficiency for the global maximum power output achieved in the far-from-equilibrium S phase and indicated by the purple closed circle is still close to the universal linear-response EMP value. This finding points out the importance of non-equilibrium phase transitions for the performance of an assembly of nano-machines and suggests synchronization as an operating mode faciliating very efficient energy-conversion processes with appreciable power output.

% I omit this supplementary discussion.
% 
%For completeness, we briefly discuss the maximization of the power output with respect to the inverse temperature illustrated in panel \ref{fig:emp}d). 
%In the SA phase, $\invtemperature < \invtemperature_c$, the maximization of the power with respect inverse temperature reads
%\begin{align}
%\frac{\D \meanfieldpower}{\D \invtemperature}  &= (\outforce - 1) \cosh \left( \frac{(\outforce - 1) \invtemperature}{2}  \right) = 0 .  \label{eq:lowbetapowerbeta}
%\end{align} 
%The $\invtemperature$-optimized equation admits the trivial solution $\outforce=1$ corresponding to equilibrium.
%
%The extremum conditions then read
%\begin{align} 
%\frac{\D \meanfieldpower}{\D \invtemperature}  &=  \euler^{-\invtemperature(1 - \outforce)} ( \outforce \!- \! 2) + \outforce = 0 . \label{eq:largebetapowerbeta}
%\end{align} 
%The maximization with respect to the temperature yields an analytic solution $\maxinvtemperature = \ln ( 2/\outforce-1 )/(1-\outforce$), which implies the trivial solution $\outforce =0 $ in compliance with the condition $\maxinvtemperature = \infty$.

\section{Conclusion and Perspectives}
\label{sec:conclusion}

%I tend not to use acronyms in intro and conclusion

We introduced and studied a thermodynamically consistent minimal model of $N$ driven and globally interacting three-state units obeying linear Markovian dynamics. 

The mean-field dynamics (which is exact when $N \to \infty$) exhibits two nonequilibrium phase transitions as a function of the inverse temperature, a Hopf and an infinite-period bifurcation. These separate three distinct phases consisting respectively of a stable fixed point where all units states are equiprobable, a limit cycle corresponding to synchronization of the units, a coexistence of three stable fixed points where the units states have unequal probabilities. 

We demonstrated that these transitions are encoded in the spectrum of the generator of the linear Markovian dynamics. The two dominant complex-conjugated eigenvalues, beside the null one, describe the mean-field dynamics over metastable times (i.e. times located between the inverse of the real parts of the next dominant eigenvalues and the inverse of their own real part) which increase with $N$. All predictions based on the spectral analysis were confirmed employing dynamic Monte Carlo simulations. 

After having established a nonequilibrium thermodynamics description of our model at different scales, we characterized the nonequilibrium phase transitions using the work dissipated by the external force driving the units. The mean-field dissipated work which reproduces very well the large $N$ results undergoes a first order phase transition followed by a second order one as a function of the inverse temperature.
When comparing a single unit to a unit in an interacting network, the average dissipated work for both units is equal in the first phase, while for the interacting unit it remarkably drops in the synchronization phase and drops even further in the third phase. Interestingly, in the presence of interactions and when $N$ is too low to produce a meaningful metastable mean-field dynamics, the average dissipated work in the second (resp. third) phase is lower (resp. higher) than for in the mean field ($N \to \infty$). 

Finally, when operating our system in the mean-field limit as a work to work converter, we found that the synchronization phase leads to a significant boost in the power output. The efficiency at maximum power of this far-from-equilibrium machine is surprisingly close to the universal linear-regime prediction.      

The model we used is minimal in that it contains the minimal ingredients to be thermodynamically consistent and at the same time give rise to a limit cycle. As most minimal stochastic models, it may find various applications (e.g. interacting molecular motors or coupled quantum dots). The methods we used are generic in that they can be used on other models.

A natural extension of this work would consist in analyzing thermodynamic fluctuations in particular close to phase transitions based on \comment{generating} function techniques and large deviation theory. 
\newcomment{Another one would be to explore the effects of local interactions and of the network topology on the dissipated work. While the qualitative behavior of synchronization is likely to survive \cite{wood2006prl,wood2006pre}, new interesting spatiotemporal regimes may emerge \cite{lindenberg2014pre}.}

At the fundamental level, our work shows an instance where bifurcation theory can be augmented with a thermodynamic interpretation to move towards a theory of nonequilibrium phase transitions. 
In such a theory, bifurcations would arise from the nonlinearities of the mean-field dynamics which emerges from an underlying stochastic thermodynamics of interacting systems in the macroscopic limit.   

From a more utilitarian perspective, our work suggest interesting avenue towards engineering interactions between assemblies of small machines to efficiently generate power, in particular in far-from-equilibrium regimes where nonequilibrium phase transitions may arise. 

%it is, in principle, possible to characterize the nonequilibrium phase transition and to determine the fluctuations in the system. These analytic predictions can then be compared with dynamic Monte-Carlo simulations or the limiting Fokker-Planck approximation of the ME \cite{herpich2018prep}. Moreover, for practical purposes it is desireable to consider a non-tightly-coupled machine exhibiting collective phenomena allowing the study of efficiency fluctuations in the vicinity of the transition between the synchronized and asynchronized.
%An obvious extension of this work would be to consider local interactions.
%In fact, the original three-state model was motivated by the feasibility of computing networks of considerable dimensions with non-global interactions.
%Even more ambitiously, investigating the dependence of the thermodynamic properties on the topology of a network is a challenging, yet intriguing problem. Indeed, there are evidences that spatiotemporal organizations can emerge for different networks with different connectivities \cite{lindenberg2014pre}. This naturally raises the question, whether these local structures - like for the global synchrony - maximize the power output and guarantee high efficiencies.

\section*{Acknowledgments}

T.H. thanks Hadrien Vroylandt and Artur Wachtel for insightful discussions. We also mention that the simulations were carried out using the HPC facilities of the University of Luxembourg \cite{hpc}. This research was supported by the National Research Fund, Luxembourg, in the frame of the AFR PhD Grant 2016, No. 11271777 and by the  European Research Council project NanoThermo (ERC-2015-CoG Agreement No. 681456).

\appendix
\section{Characterization of Hopf bifurcation}
\label{sec:hopfbifurcationproof}

We shall in the following prove that the Hopf bifurcation observed in Sec. \ref{sec:meanfielddynamics} is supercritical, \idest results in stable LCs. To characterize the LC close to the bifurcation point, we consider the normal form of the Hopf bifurcation. The procedure is detailed in \cite{kuznetsov1998}.

At first, we transform the two-dimensional system in Eq. (\ref{eq:meanfieldmasterequation}) into a single equation
\begin{align} \label{eq:singlecomplexequation}
\dot{\complexvar} &= \meanfieldeigenvalue(\Delta \invtemperature) \complexvar + g(\complexvar, \complexvar^{*},\Delta \invtemperature) ,
\end{align}
where $\complexvar$ is a complex variable, $\complexvar^{*}$ its complex-conjugate, $\Delta \invtemperature=\invtemperature-\invtemperature_{c_1}$ gives the distance of the inverse temperature to the critical inverse temperature of the Hopf bifurcation and $g=O \big(\begin{Vmatrix} \complexvar \end{Vmatrix}\!\textsuperscript{2} \big)$ is a smooth function of $(z,z^{*},\Delta \invtemperature)$.

Such a transformation is achieved by first finding the complex eigenvectors $\bm{r}$ and $\bm{v}$ determined by
\begin{align}
\jacobian(0)\bm{r} = \meanfieldeigenvalue(0) \bm{r}, \quad \jacobian(0)^{\top}\bm{v} = \meanfieldeigenvalue(0)^{*} \bm{v} ,
\end{align}
where the real and non-symmetric Jacobian $\jacobian$ resulting from the linearization of Eq. \eqref{eq:meanfieldmasterequation} is evaluated at the bifurcation point $\invtemperature = \invtemperature_{c_1}$, yielding
\begin{align}
\bm{r} &= \left( \frac{1}{2} (- 1 + \sqrt{3} \, \i) , 1  \right)^{\top} \\
\bm{v} &= \frac{1}{3-\sqrt{3}\, \i} \left( 1+\sqrt{3} \, \i , 2   \right)^{\top} .
\end{align}
If $|\Delta \invtemperature|$ is sufficiently small, the two-dimensional system from Eq. (\ref{eq:meanfieldmasterequation}) can be written as 
\begin{align} \label{eq:taylormeanfieldmasterequation}
\dot{\meanfieldprobabilityvector} = \jacobian(\Delta \invtemperature) \meanfieldprobabilityvector + \bm{F}(\meanfieldprobabilityvector, \Delta \invtemperature) ,
\end{align}
where $\bm{F}(\meanfieldprobabilityvector, \Delta \invtemperature ) $ is a smooth vector function whose components have Taylor expansions in $\meanfieldprobabilityvector$ starting with at least quadratic terms, $F_{1,2}$=$O \big( \begin{Vmatrix} \meanfieldprobabilityvector \end{Vmatrix}\! \textsuperscript{2} \big) $. Using Eq. \eqref{eq:singlecomplexequation} and the properties $\langle \bm{v}, \bm{r} \rangle = 1$,$\langle \bm{v}, \bm{r}^* \rangle = 0$, one can show that 
{\small  \begin{align}
\!\! g(\complexvar, \complexvar^{*}\!\!,\Delta \invtemperature) \!=\! \langle \bm{v} (\Delta \invtemperature), \bm{F}(\complexvar \, \bm{r} (\Delta \invtemperature) \!+\! \complexvar^{*}  \bm{r}^{*}(\Delta \invtemperature), \Delta \invtemperature ) \rangle .
\end{align} }
The function $g$ can be formally written as a Taylor series in the two complex variables $\complexvar$ and $\complexvar^{*}$,
\begin{align}
& g( \complexvar,\complexvar^{*},\Delta \invtemperature ) = 
\sum\limits_{k+l \geq 2} \frac{1}{k! l!} 
\frac{\D^{k+l}}{\D z^k \D z^{*^l} } 
g_{kl}(\Delta \invtemperature)
 \, z^k z^{*^l}  ,
\intertext{with}
\! &g_{kl}(\Delta \invtemperature) \!\! = \!\!\!
 \left. \langle \bm{v} (\Delta \invtemperature), \!\bm{F}(\! \complexvar \bm{r} (\Delta \invtemperature) \!+\! \complexvar^{\! *}  \bm{r}^{\! *}(\Delta \invtemperature), \Delta \invtemperature ) \rangle \right|_{z=0} \! .
\end{align}

Moreover, if the function $F(\meanfieldprobabilityvector,\Delta \invtemperature)$ from Eq. (\ref{eq:taylormeanfieldmasterequation}) is represented as
\begin{align}
F(\x,0) = \frac{1}{2} B(\x,\x) + \frac{1}{6} C(\x,\x,\x)+ O\left( \begin{Vmatrix} x \end{Vmatrix}^4 \right) ,
\end{align}
where  $ B(\x,\y)$ and $ C(\x,\y,\bm{u}) $ are \textsl{symmetric} multilinear vector functions of $\x,\y,\bm{u} \in \mathbb{R}^2$, it follows that
\begin{subequations}
\begin{align}
g_{20} &= \langle \bm{v}, B(\bm{r},\bm{r}) \rangle =0 \\ 
g_{11} &= \langle \bm{v}, B(\bm{r},\bm{r}^{*} ) \rangle =0 \\
g_{21} &= \langle \bm{v}, C(\bm{r},\bm{r},\bm{r}) \rangle .
\end{align}
\end{subequations}
In coordinates, one has for these vector functions
\begin{align}
B_i(\x,\y) &= \sum\limits_{j,k=1}^2 \left. \frac{\D^2 F_i(\bm{\xi},0)}{\D \xi_j \D \xi_k} \right|_{\xi =0} \!\!\! x_j \, y_k, \;\; i\!=\!1,2 \\
C_i(\x,\y,\bm{u}) &= \!\! \sum\limits_{j,k,l=1}^2 \left. \frac{\D^3 F_i(\bm{\xi},0)}{\D \xi_j \D \xi_k \D \xi_l} \right|_{\xi =0} \!\!\!\!\! x_j y_k u_l, \; i\!=\!1,2 .
\end{align}

With these expressions at hand, we can determine the first Lyapunov coefficient $L_1$ as
\begin{align} \label{eq:lyapunov}
L_1 = \frac{1}{2 \, \limitcyclefrequency^2} \, \Real \left( \i \, g_{20} \, g_{11} + \limitcyclefrequency \, g_{21} \right) ,
\end{align}
where the eigenvalue of the Jacobian is decomposed as $\lambda(\Delta \invtemperature)$=$\sigma(\Delta \invtemperature)$+$\i \, \omega(\Delta \invtemperature)$ and
\begin{align} 
\limitcyclefrequency = \left. \meanfieldeigenvalue(\Delta \invtemperature) \right|_{\invtemperature = \invtemperature_{c_1}} = \arrheniusprefactor \sqrt{3} \, \sinh \left( - \frac{3 \force}{2 \potential} \right)  \label{eq:eigenfrequency}
\end{align}
is the LC frequency, $\limitcyclefrequency \equiv \omega(0) $, evaluated at the bifurcation point $\Delta \invtemperature$=0.
For Eq. (\ref{eq:lyapunov}) to hold, the two requirements $\omega(0) > 0$ and $ \sigma^{\prime} (0) < 0$ must be met.
From Eq. (\ref{eq:eigenfrequency}) and
\begin{align}
\sigma^{\prime} (0) = \potential \, \arrheniusprefactor \, \cosh \left( \frac{3 \force}{2 \potential} \right) \,
\end{align}
it follows that this is only true for attractive interactions, $\potential <0$. Collecting results, we finally arrive at 
\begin{align} \label{eq:lyapunovresult}
L_1 = -\frac{81}{2} \arrheniusprefactor  \cosh \left( \frac{3 \force}{2 \potential} \right) ,
\end{align}
which is negative for any $\potential <0$, hence for attractive interactions stable LCs emerge at the bifurcation point $\invtemperature_{c_1}$ as asserted above.

\section{Equal-probability of stationary microstates belonging to a macrostate}
\label{sec:spanningtreeproof}

A special case for which also the EP and system entropy can be exactly represented by macrostate ensemble quantities is the nonequilibrium steady state reached at large times. The probabilities associated with states in the stationary regime can be calculated via the spanning tree formula. We denote the graph representing the network by $\graph$. A spanning tree, $\tree'(G)$ of a graph is defined as a covering subgraph of $\graph$, \idest all of its edges are also edges of $G$ and it contains all vertices (microstates) of $G$. It is furthermore required that $\tree'(\graph)$ is connected and contains no circuits. We introduce the notation $\mathcal{A}(\tree'^{(\mu)}_{\microstate}(\graph))$ referring to the $\mu$th spanning tree rooted in $\microstate$, that is a tree whose branches are pointing towards the vertex $\microstate$.
The spanning tree formula states \cite{schnakenberg1976rmp}
\begin{align} \label{eq:spanningtreeformula}
\microsteadyprobability_{\microstate} \!\!=\! 
\frac{\sum\limits_{\mu} \mathcal{A}(\tree'^{(\mu)}_{\microstate}(\graph)) }{\sum\limits_{\microstate} \sum\limits_{\mu} \mathcal{A}(\tree'^{(\mu)}_{\microstate}(\graph))} \!=\!
\frac{ \sum\limits_{\tree'_{\microstate}(\graph)} \prod\limits_{ \substack{ \text{s.t. current} \\ \text{ is directed to } \microstate }} \microrates_{\microstate'\microstate''} }{ \sum\limits_{\microstate}\sum\limits_{\tree'_{\microstate}(\graph)} \prod\limits_{ \substack{ \text{s.t. current} \\ \text{ is directed to } \microstate }} \microrates_{\microstate'\microstate''} } \, .
\end{align}
As was already discussed above, the transition rates do not depend on the microstates belonging to the same pair of macrostate. Moreover, the connectivity of the network is also not a function of the microstate, since, due to the all-to-all interaction, the number of edges of any vertex in the microspace network is always $2 \dimension$, such that the number of spanning trees rooted in $\microstate$ is constant for all $\microstate$ inside the same macrostate. Thus, at steady state, all microstates constituting the same macrostate
\begin{align}
\microsteadyprobability_{\microstate} \! &= \! \frac{ \!\!\!\! \sum\limits_{\tree'_{\microstate \in \macrostate}(\graph)} \prod\limits_{ \substack{ \text{s.t. current}  \\ \text{ is directed to } \microstate \in \macrostate }} 
\left. \microrates_{\microstate \microstate'} \right|_{\substack{\microstate \in \macrostate \\ \microstate' \in \macrostate'} }  }{ \sum\limits_{\microstate \in \macrostate} \sum\limits_{\tree'_{\microstate(\dimension)}(\graph)} \!\!  \prod\limits_{ \substack{ \text{s.t. current} \\ \text{ is directed to } \microstate \in \macrostate }} \!\!\!\!\!\! \!\!\!\!  
\left. \microrates_{\microstate \microstate'} \right|_{\substack{\microstate \in \macrostate \\ \microstate' \in \macrostate'}}
 } \!=\! \mathrm{const} ,
\end{align}
are equally probable and hence
\begin{align} \label{eq:steadystateequalmicroprobabilities}
\microsteadyprobability_{\microstate} &= \frac{\macrosteadyprobability_{\macrostate}}{\multiplicity(\macrostate)} \, ,
\end{align}
where $\multiplicity(\macrostate)$ is the number of microstates forming the macrostate $\macrostate$ given by a trinomial coefficient of the occupation numbers $\occupation_i$ determined in Eq. (\ref{eq:multiplicityfactortrinomial}).

\section{Stationary solution for single unit}
\label{sec:singleunitproof}

We consider a single unit with states $i=1,2,3$ whose evolution is governed by the ME
\begin{align}
\macroprobability_i = \sum\limits_{i,j} \macrorates_{ij} \, \macroprobability_j \, ,
\end{align}
where $\macroprobability$ is the (macro-)probability to find the unit in the single state $i$ with the transition rates
\begin{align}
\macrorates_{ij} = \euler^{ - \frac{\invtemperature}{2} \left( \stateenergy_i - \stateenergy_j + \meanfieldsign \, \force  \right) } \, , 
\end{align}
with the sign function $\meanfieldsign$ as defined in Eq. (\ref{eq:meanfieldlocaldetailedbalance}) ensuring the validity of local detailed balance.
The steady-state work current reads
\begin{align} 
\langle\dot{\macrowork}^s \rangle &= f \, \sum\limits_{i,j}   \meanfieldsign \, \macrorates_{ij} \, \macroprobability^s_{\j} \, .
\end{align}
Using the spanning tree formula from Eq. (\ref{eq:spanningtreeformula}), one obtains for the stationary probabilities
\begin{align}
\macroprobability_1^s &= \frac{ a_1 }{ a_1 + a_2 + a_3 } , \quad \macroprobability_2^s = \frac{ a_2 }{ a_1 + a_2 + a_3 } \, ,
\end{align}
where
\begin{subequations}
\begin{align}
a_1 &= W_{13} W_{12} \!+\! W_{12} W_{23} \!+\! W_{13} W_{32} \\
a_2 &= W_{23} W_{31} \!+\! W_{21} W_{13} \!+\! W_{23} W_{21} \\
a_3 &= W_{31} W_{12} \!+\! W_{32} W_{21} \!+\! W_{32} W_{31} ,
\end{align}
\end{subequations}
For a flat energy landscape, $\stateenergy_i $=$ \const$, we indeed find that the symmetric stationary solution $\macroprobability_i = 1/3$ is independent of $\invtemperature$ and $\force$ like in the MF limit. Next, the stationary work current is given by \vspace*{-1cm}

\begin{widetext}
\begin{align}  \label{eq:singleunitworkcurrent}
\langle\dot{\macrowork}^s \rangle &= 3f \, \frac{ W_{13} W_{21} W_{32} - W_{31} W_{12} W_{23} }{ W_{12}(W_{13}+W_{23}+W_{31}) 
+ W_{13}(W_{21}+W_{32})
+ (W_{21}+W_{31})(W_{13}+W_{23}+W_{31})} 
\end{align}
\end{widetext} \vspace*{-1cm}

that simplifies to $\langle \dot{\mathcal{\macrowork}}^{s} \rangle $=$2 \, \arrheniusprefactor \force \sinh \left( \force \invtemperature /2 \right) $ (see Sec. \ref{sec:dissipatedwork}).

\bibliography{bibliography.bib}

\end{document}